\documentclass[11pt,a4paper]{article}

\usepackage[utf8]{inputenc}
\usepackage[T1]{fontenc}
\usepackage{lmodern}
\usepackage[margin=25mm]{geometry}
\usepackage{microtype}
\usepackage{amsmath,amssymb}
\usepackage{graphicx}
\usepackage{booktabs}
\usepackage[font=small,labelfont=bf]{caption}
\usepackage[sort&compress,numbers]{natbib}
\usepackage{fancyhdr}
\usepackage{xcolor}
\usepackage[hyphens]{url}
\usepackage[colorlinks=true,allcolors=blue!60!black]{hyperref}
\hypersetup{
  pdftitle={Energy harvesting ambient vibration},
  pdfauthor={Igor A. Khovanov},
  pdfkeywords={Energy harvesting; Piezoelectricity; Nonlinear effects; Bistability; Stochastic perturbations; Harmonic noise}
}
\title{\bfseries\Large Energy harvesting ambient vibration}
\author{Igor A. Khovanov\thanks{\href{mailto:i.khovanov@warwick.ac.uk}{i.khovanov@warwick.ac.uk};
  ORCID: \href{https://orcid.org/0000-0002-0372-8457}{0000-0002-0372-8457}.
  This is an original manuscript of an article published by Taylor \& Francis in \emph{Contemporary Physics}
  on 24 August 2026 under the title ``Piezoelectric energy harvesting from ambient vibrations'',
  available at: \url{https://doi.org/10.1080/00107514.2026.2716476}.}\\[6pt]
  \normalsize\itshape School of Engineering, University of Warwick,\\
  \normalsize\itshape Coventry CV4 7AL, United Kingdom}
\date{}

\begin{document}

\maketitle
\thispagestyle{plain}

\begin{abstract}
\noindent
Converting the mechanical power of ambient environmental vibrations into electrical form allows us to avoid batteries for remote sensors. Understanding and describing the properties of ambient vibration is the first, often-missed step toward efficient power conversion. We consider examples of machinery and vehicle vibrations to demonstrate how their properties can be characterised and to illustrate the key features of ambient vibrations. Another step is to design the corresponding harvester. A piezoelectric harvester consisting of a cantilever that converts ambient vibration into cantilever oscillations and a piezoelectric layer that converts the oscillations into electric charge is considered. The importance of the cantilever's nonlinearity to harvester performance is discussed. Then, by considering several signals representing ambient vibration, it was shown that the interaction between ambient vibration features and the harvester's nonlinearity affects harvester performance. This interaction can degrade or improve the harvester's conversion efficiency. Some open questions related to exploring the nonlinearity for maximising harvester efficiency and to developing adequate models of ambient vibration are outlined.

\end{abstract}

\section{Introduction}

Around three decades ago, the development of ultra-low-power microelectronics and wireless sensors created the need to power sensors located remotely or in hard-to-access places \cite{Priya:09}. Batteries can address the need, but this solution is impractical, since it either assumes long-lasting, expensive batteries or frequent replacement with cheaper ones. Therefore, alternative options were explored, including scavenging ambient vibration to generate usable electrical power for microelectronic devices.  The required power is typically very low, from tens of microwatts to a few milliwatts, and often needs to be delivered for a short period of time.  Devices scavenging mechanical energy from vibration are called harvesters, and they aim to provide enough energy to run, for example, a wireless sensor or a medical implant without batteries. Currently, demand for harvesters is growing as the spread of self-powered Internet of Things and "deploy-and-forget" sensors.  

The term { \it ambient} assumes that ambient vibration provides free, leftover energy from the environment that would otherwise go to waste. Simultaneously, it means that ambient vibration is provided as it is, with properties that are hardly controllable and amendable. So, harvesters should be designed to maximise the power output for given vibration properties. Apparently, this maximisation task is a challenge which often remains untackled. Below, we will discuss it in more detail.

The other two important tasks are converting mechanical power to electrical power and utilising the latter for micro-electronic devices. Different conversion principles, mainly based on piezoelectric, electromagnetic, electrostatic and triboelectric transduction, are successfully employed \cite{Beeby:11}. In piezoelectric conversion, bending strains the material, releasing electrons and generating voltage. This conversion is attractive for its high power density, relative ease of implementation in a harvester, and status as the de facto ``textbook'' harvester. A current induced by a magnet moving relative to a coil is the basis of electromagnetic conversion, which is a simple and reliable option. Still, it is less efficient and bulkier than the piezoelectric one. Electrostatic transduction utilises a variable capacitor whose capacitance changes with vibration and requires an initial charge to be applied to the capacitor. Triboelectric harvesters employ static electricity generated by friction between two different materials in contact, which induces charges, and the subsequent relative movement drives electrons between the electrodes. Mechanical interactions between materials reduce the durability of triboelectric conversion, but it could be extremely efficient for slow vibrations, such as those observed during human motion. 

The electricity generated by harvesters typically reflects the properties of ambient vibration, and it is an oscillatory voltage (current) that requires adaptation for sensor purposes, or sensors should be designed to accommodate the oscillatory nature of electrical energy.  The simplest solution is to rectify the alternating voltage, then adjust the charging voltage applied to a rechargeable battery, or provide a constant polarity to sensors. Many different power-management electrical circuits have been proposed, and this area is an active area of research \cite{Szarka:2018}.

Thus, harvesting mechanical energy involves several distinct tasks and steps that are usually considered independently across different engineering branches. In this paper, we focus on the first step, namely designing a harvester to maximise energy scavenging from ambient vibration. Initially, some examples of ambient vibration are considered, and their features are discussed. Then, a typical piezoelectric harvester and its mathematical model are introduced. Further, the design task is considered in terms of mathematical predictive modelling of ambient vibration and harvesters. Different models of ambient vibration are introduced. For each model, the corresponding adjustments to the harvester to maximise performance are considered. The consideration is linked to several physical phenomena that are typically utilised to maximise harvester output. A brief conclusion summarises the current state of the task of maximising the efficiency of energy harvesting from ambient vibration.
 
 \section{ Ambient vibration }

Sources of ambient vibration are countless. Each source has its unique features and energy flow. Our natural environment produces seismic ground motion and wind- and flow-induced vibration of objects. Civil structures such as bridges, railway tracks, buildings, walls, and floors move due to structure-specific causes, such as traffic, wind, or footfall. Rotating and reciprocating machinery, including motors, pumps, compressors, fans, and household appliances, is designed to perform a wide range of movements. Vibration persists during transportation and is observed in various parts of a vehicle, such as the chassis, engine compartments, aircraft fuselage, ship hulls, and rolling stock.  Vibrations of different parts of the body are linked to biological motion, for example, when we walk, run, breathe, and heartbeat. In all these examples, vibration is associated with a body that moves by changing its geometry, location, or both.  Causes of vibration can be internal or external to the body, depending on whether they involve interaction with the surroundings or with interacting parts.  So, the vibration's origin has many facets, and a comprehensive, unique framework for describing vibration remains an open question.  Nonetheless, we need tools to compare different vibration sources and, more importantly, to select features (properties) that are important for designing energy harvesters. 
 
 Vibration assumes that a body or part of it moves in a particular way, namely, back and forth about a body's (or its part's) position or positions. These positions correspond to equilibrium states, in which the body or its parts should be if the movement's causes are removed. One can say that a body oscillates around its state(s), and the oscillation has a mechanical nature. The recurrent character of vibration is an important component of the motion, since by alternating its position, and, consequently, its velocity, the body ``produces'' acceleration, and, therefore, a mechanical force which we can scavenge. Thus, the ambient vibration characterisation should include information about the force or the body's acceleration. This information gives us an idea about how much energy is available.

Additionally, we need some tools to describe properties of the observed back-and-forth motion. Also, we can simplify our analysis of ambient vibration sources by treating the vibrating object as a lumped system. This simplification is a widely used technique in physics and engineering modelling that considers an entire object or a part of it as a single point. The latter means that we ignore the geometry of the objects and focus on a point source of a force.

Let us consider specific examples of ambient vibration sources to illustrate the vibration properties important for energy harvesting and the descriptive tools available to characterise them. These examples do not directly relate to energy harvesting and were considered to characterise the vibration of different objects using high-quality measurements. The latter is a dominant factor for selecting the examples.

  \subsection{ Machinery vibration }
 
 The considered data set is associated with the paper \cite{Qiu:06}, in which the authors use a test ring to prognosticate bearing failure. In the ring, a motor rotates a shaft at a constant speed of 2000 revolutions per minute, equivalent to 33.3(3) Hz. The shaft is secured by the inner rings (often called races) of four bearings; the outer ring of which is seated in a housing (pedestal block) fixed to the loaded test frame. Two rows of rollers are located between the rings. The accelerometer is bolted to the housing, so it senses the housing's motion in radial (perpendicular to the shaft) direction. In the measurements, a seismic accelerometer (PCB 353B33 ICP) that reports absolute acceleration relative to the inertial frame was used. Short, around 1-second recordings were sampled at $f_s=20$ kHz rate every 20 minutes over a long period (more than a month) until bearing failure. A significant part of this period, the bearings were healthy, and that part is discussed in this subsection. 
 
 \begin{figure}[!tb]
\centering
	\includegraphics[width=0.48\columnwidth]{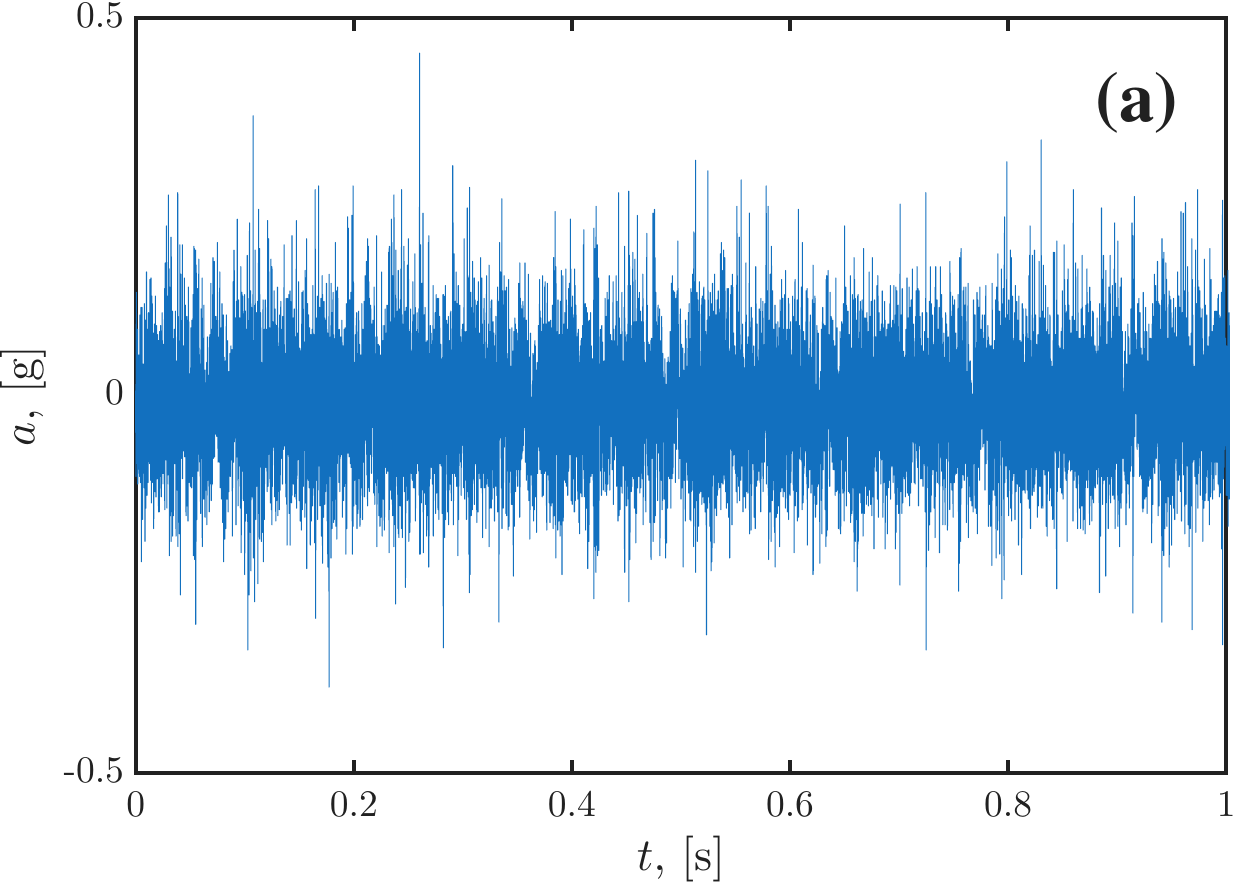} ~~
	\includegraphics[width=0.48\columnwidth]{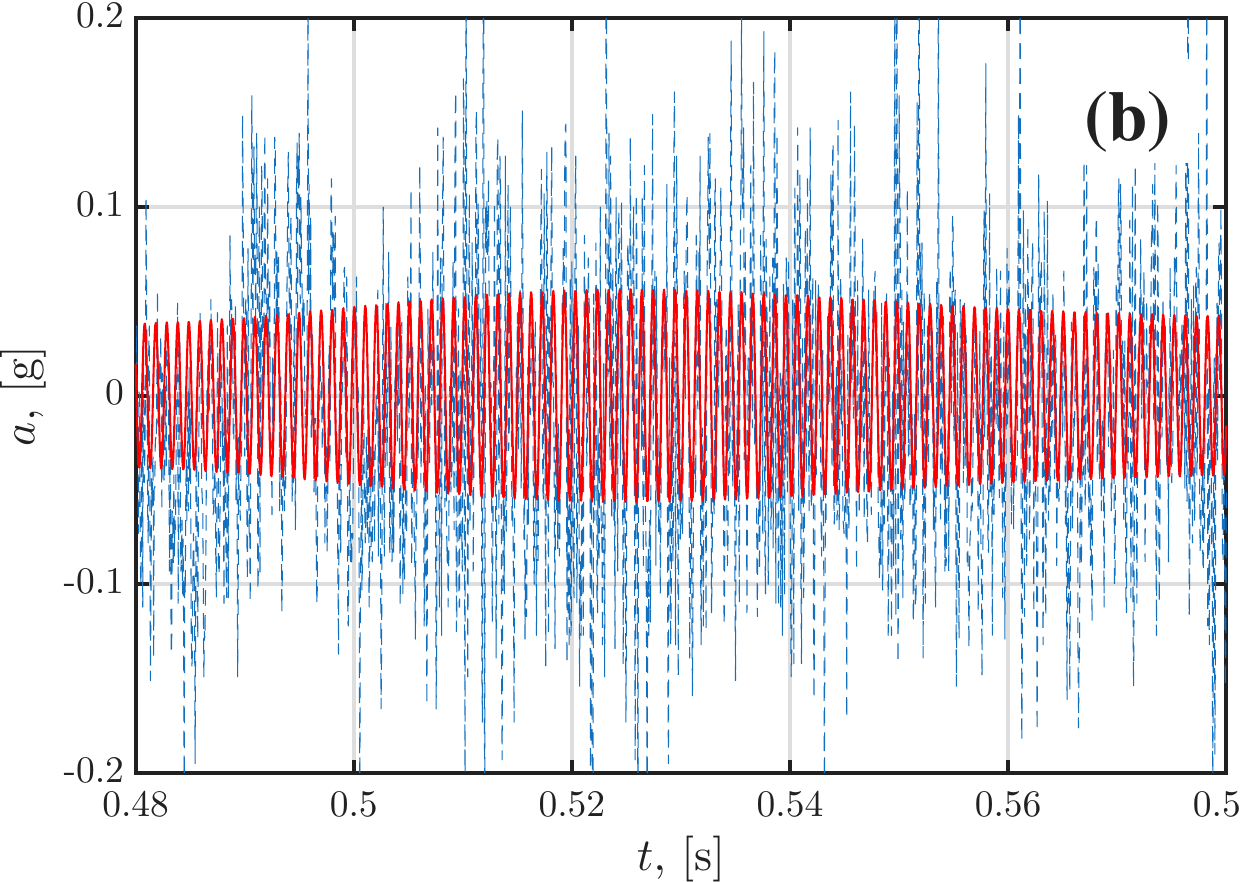}  
\caption{(a) An example of the accelerometer output, a time trace of $a(t)$, in a machinery vibration test. (b)  A short part of the signal from Figure (a) is shown with the dashed blue line. The solid red line shows the filtered signal, which contains frequencies around a dominant spectral peak (see the text for further detail). } \label{figure1}
\end{figure}
 
\begin{figure}[!tb]
\centering
	\includegraphics[width=0.48\columnwidth]{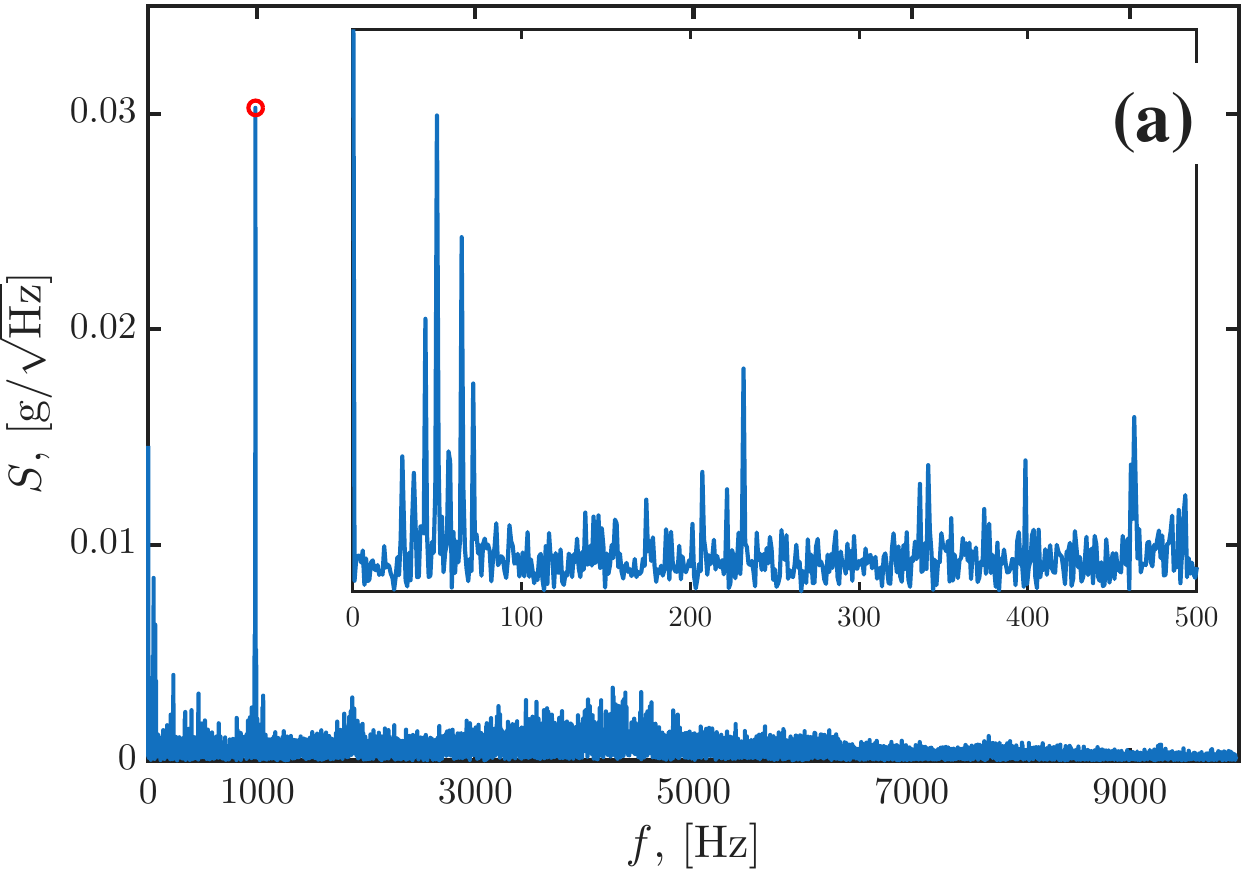} ~~
	\includegraphics[width=0.48\columnwidth]{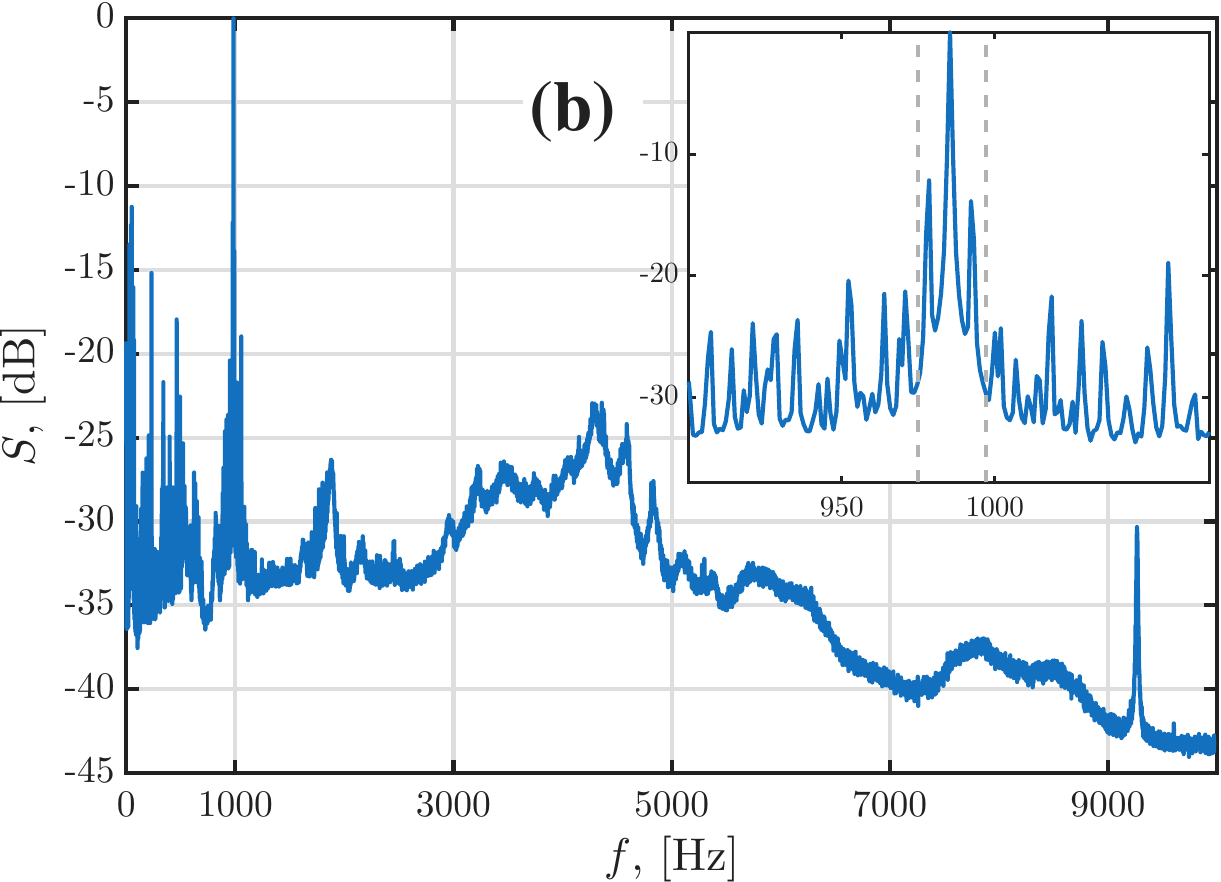}  	
\caption{(a) The amplitude spectrum $S(f)$ of a single time trace $a(t)$ shown in Fig.\,\ref{figure1}\,(a). Note, the linear scale is used for the y-axis. The inset shows a low-frequency part of $S(f)$. The marker $\circ$ denotes the dominant single harmonic component. (b) The amplitude spectrum $S(f)$ averaged over 300 time traces of $a(t)$. The decibel units are used for $S(f)$. The inset shows $S(f)$ around a dominant harmonic. The dashed vertical lines denote the frequency range to estimate the filtered signal shown in Fig.\,\ref{figure1}\,(b) with the solid red line. } \label{figure2}
\end{figure}

 An example of an accelerometer's recording, $a(t)$, is shown in Fig.\,\ref{figure1}\,(a). The acceleration was measured in units of g ($\text{g}= 9.80665 \text{ m}/\text{s}^2$, standard acceleration due to the Earth's gravity. The signal $a(t)$ is irregular and has no visible pattern. Such signals are usually associated with noise. The amplitude spectrum $S(f)$ of the signal gives its frequency content: the amplitude $S$ at a given frequency $f$ is evaluated using the fast Fourier transform. The value of $S$ is the root mean square amplitude of the power spectral density, which is typically used in engineering to preserve the signal energy in its spectral representation. Therefore, the units for $S$ are g/$\sqrt{ \textrm{Hz} }$. The acceleration values are sampled, so the signal is defined in discrete time instances $t_i=i/f_s$, where $i=1, 2, 3 \ldots$. Therefore, applying the fast Fourier transform leads to the finite set of frequencies $f_k$, which range from zero until the Nyquist frequency $f_s/2$ (see, for example, the book \cite{bendat_random_2011} for further details). As a result, periodic components appear as large, narrow, visible peaks, and noise appears as erratic peaks forming a background in the spectrum. Figure \ref{figure2}\,(a) shows the amplitude spectrum of the acceleration signal depicted in Fig.\,\ref{figure1}\,(a). There is one large peak, denoted with marker $\circ$ in the figure and located at $f \approx 985$ Hz. Also, there are many smaller peaks around this dominant peak and in the low frequency range (see the inset in Fig.\,\ref{figure2}\,(a)). A noisy background spreads across all frequencies. 

Similar to the signal $a(t)$ shown in Fig.\,\ref{figure1}\,(a), the spectrum $S(f)$ in Fig.\,\ref{figure2}\,(a) is also random. It means the number of peaks and their locations, as well as the amplitude of the noisy background, vary between recorded signals. More statistically stable information can be obtained through averaging. The procedure consists in evaluating power spectra $S_j^2(f)$ for $N$ individual recordings and their averaging: $S(f)=\sqrt{ \frac{1}{N} \sum_{j=1}^{N} S_j^2(f)}$. As a result, the erratic changes in noise components became smoother, and the periodic components are more visible. The amplitude spectrum obtained by averaging the power spectra of $N=300$ recordings is shown in Fig.\,\ref{figure2}\,(b) using the decibel, dB, scale: $20 \log_{10} (S(f)/S_{max})$, where $S_{max}$ is the maximum in $S(f)$. The logarithmic scale allows us to visualise spectral contributions that differ by orders of magnitude. For example, a 6 dB difference corresponds to a factor of 2, and a 20 dB difference corresponds to a factor of 10. Such changes are hardly seen in the linear scale used in Fig.\,\ref{figure2}\,(a). 

The average spectrum (Fig.\,\ref{figure2}\,(b)) confirms that the peak at $f \approx 985$ Hz is dominant. Two peaks around it on the left and right sides (within the dashed lines in the inset in Fig.\,\ref{figure2}\,(b)) indicate that the amplitude of the harmonic signal at $f \approx 985$ Hz is modulated. Note that other smaller peaks contribute to the modulations, but their contribution is considerably weaker. 

We can filter out the portion of the signal associated with the dominant peak using a Fourier-based filter. The procedure is the following. Consider a recording, apply the fast Fourier transform, put all transform coefficients but those around the peak (within the dashed lines in the inset of Fig.\,\ref{figure2}\,(b)) equal to a small value, $10^{-32}$, and perform the inverse transform. The resulting signal is shown in Fig.\,\ref{figure1}\,(a) by the solid red line. The slow time-varying amplitude of the filtered signal is visible. Also, the amplitude is smaller than that of the unfiltered signal (the dashed line). 

Another distinguishing feature of the spectrum shown in Fig.\,\ref{figure2}\,(b) is a strong noise background in high frequencies. The spectrum can be decomposed into three parts: a low-frequency component with a complex mixture of periodic components (13\% of the power), a dominant peak with two side peaks (17\% of the power), and high-frequency noise (70\% of the power). Thus, the dominant peak accounts for only a relatively small proportion of the power, and wide-band noise is a dominant component of the signal. 

\begin{figure}[!tb]
\centering
	\includegraphics[width=0.48\columnwidth]{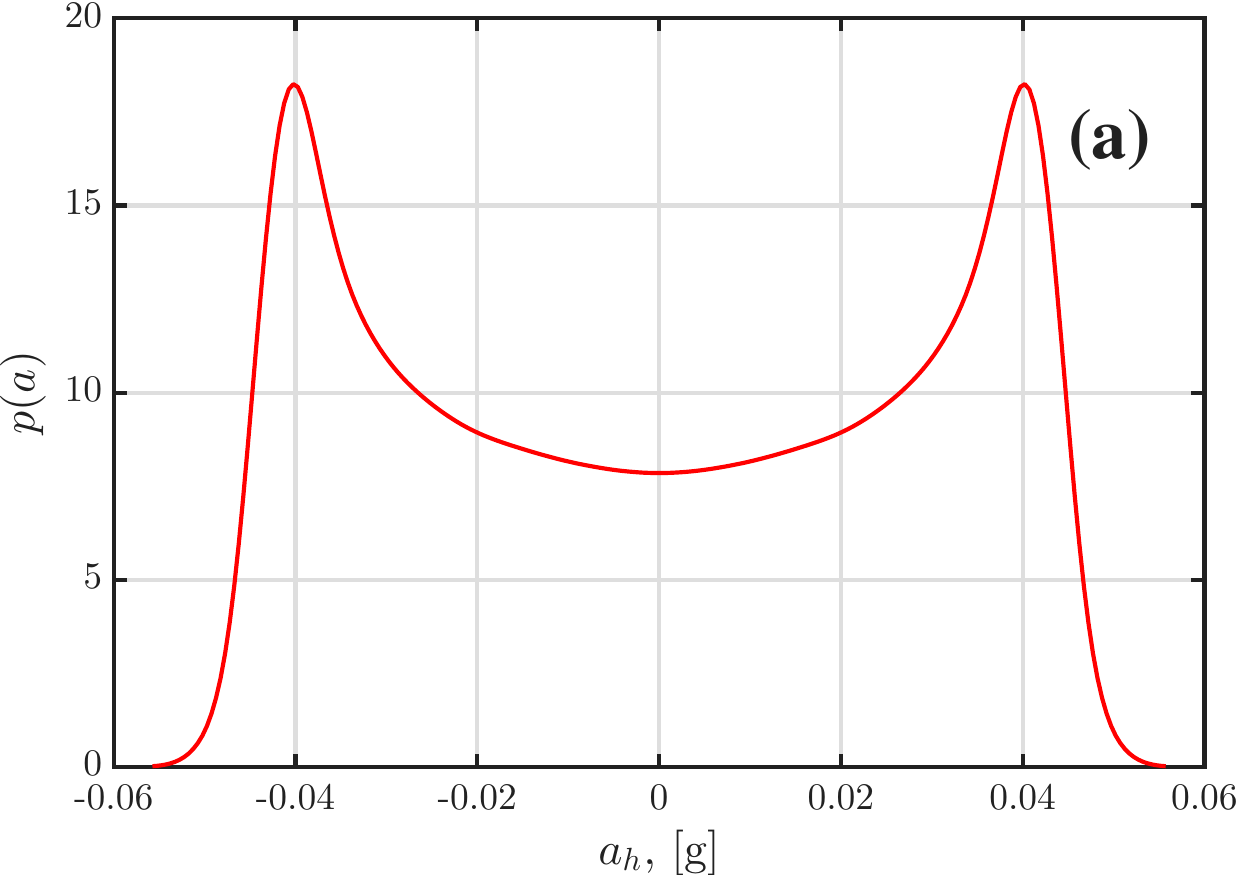} ~~
	\includegraphics[width=0.48\columnwidth]{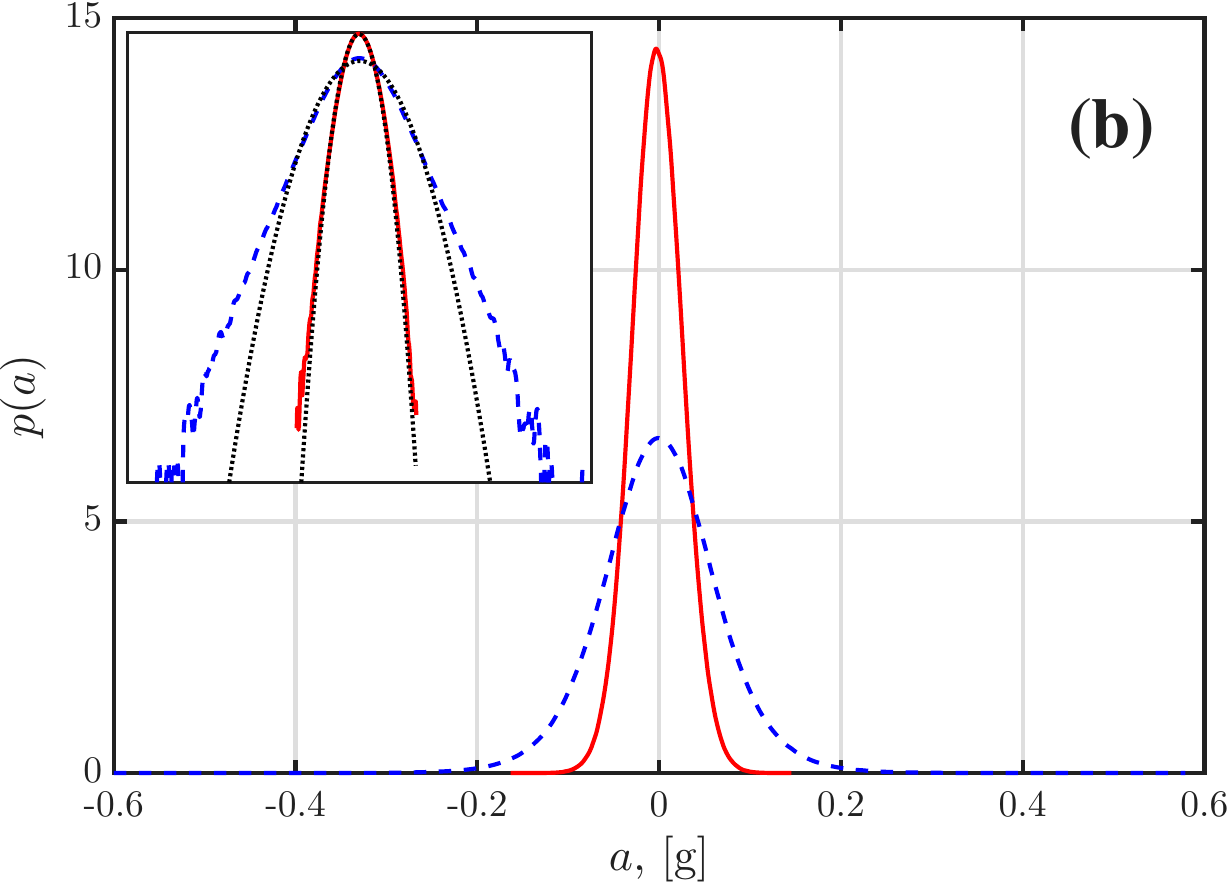}  	
\caption{ (a) The probability density function $p(a)$ for the harmonic component  $a_{h}(t)$. (b)  The probability density functions $p(a)$ for low-frequency $a_{lf}(t)$ and high-frequency  $a_{hf}(t)$ components are shown with the solid red and dashed blue lines, respectively. The inset shows both functions $p(a)$ on a logarithmic y-axis. In the inset, the x-axis range is the same as in the main figure. The dotted black lines show the fitted normal probability density function.s } \label{figure3}
\end{figure}

So far, to gain insight into the structure of vibration, we have used spectral analysis. However, it gives a part of the story for a random signal. For complete characterisation, as in the book \cite{bendat_random_2011}, we need to consider the distribution of the signal. Let's use the filter on the base inverse Fourier transform and decompose each recording, $a(t)$, into three components: the low frequency part $a_{lf}(t)$ (from zero frequency to the first dashed line in the inset of Fig. \ref{figure2}\,(b)), the harmonic signal $a_h (t)$, associated with the dominant peak (the frequency range between the dashed line), and the high frequency part $a_{hf} (t)$ (the frequency range from the second dashed line until Nyquist frequency). The probability density functions for the components are shown in Figure \ref{figure3}. The distributions were obtained for the same 300 recordings used for the spectral density estimation. The distribution for $a_h (t)$ (Fig. ~\ref{figure3}\,(a)) is close to that for the sine function, but with small tails linked to the modulation and weak noise contributions. The distributions for $a_{lf} (t)$ and $a_{hf} (t)$ (Fig. ~\ref{figure3}\,(b)) are different. They are unimodal and close to the Gaussian (normal) distribution. However, the Kolmogorov-Smirnov test rejects the null hypothesis that both components follow normal distributions. The tails of the distributions are different as it is illustrated by the dotted lines in the inset of Fig. \ref{figure3}\,(b). 

Thus, in this subsection, we consider an example of ambient vibration and apply two tools: spectrum and distribution, to characterise the vibration properties. The tools allow us to conclude that a seemingly random signal has a complex structure, comprising several components. One component is close to a harmonic signal and shows a strong spectral peak. A low-frequency component consists of a mixture of many periodic contributions that result in a unimodal distribution distinct from those for the harmonic signal. Finally, there is a high-frequency noisy component with a wide-band spectrum and an unimodal distribution distinct from normal. This high-frequency noise contains the most power of the ambient vibration.

 \subsection{ Vehicle vibration }

\begin{figure}[!tb]
\centering
	\includegraphics[width=0.48\columnwidth]{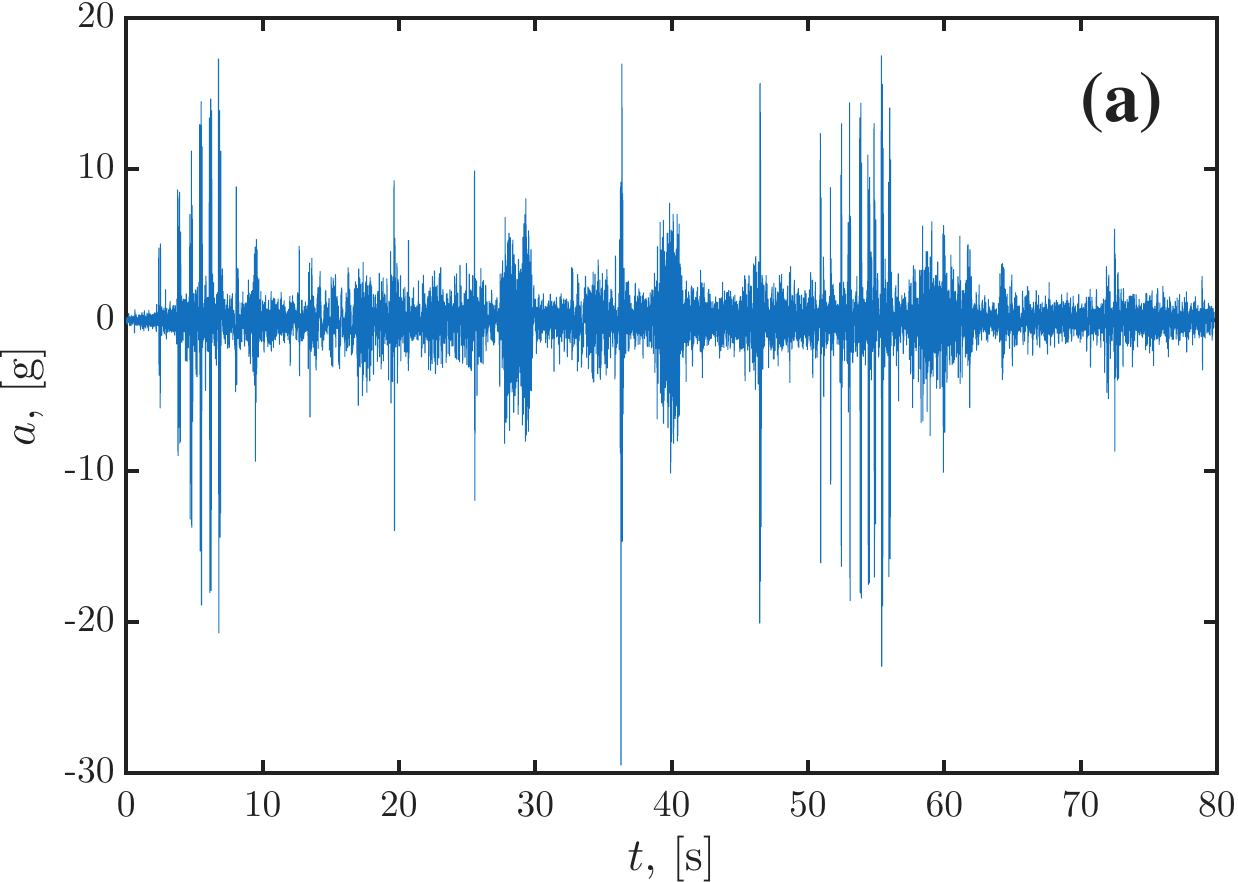} ~~
	\includegraphics[width=0.48\columnwidth]{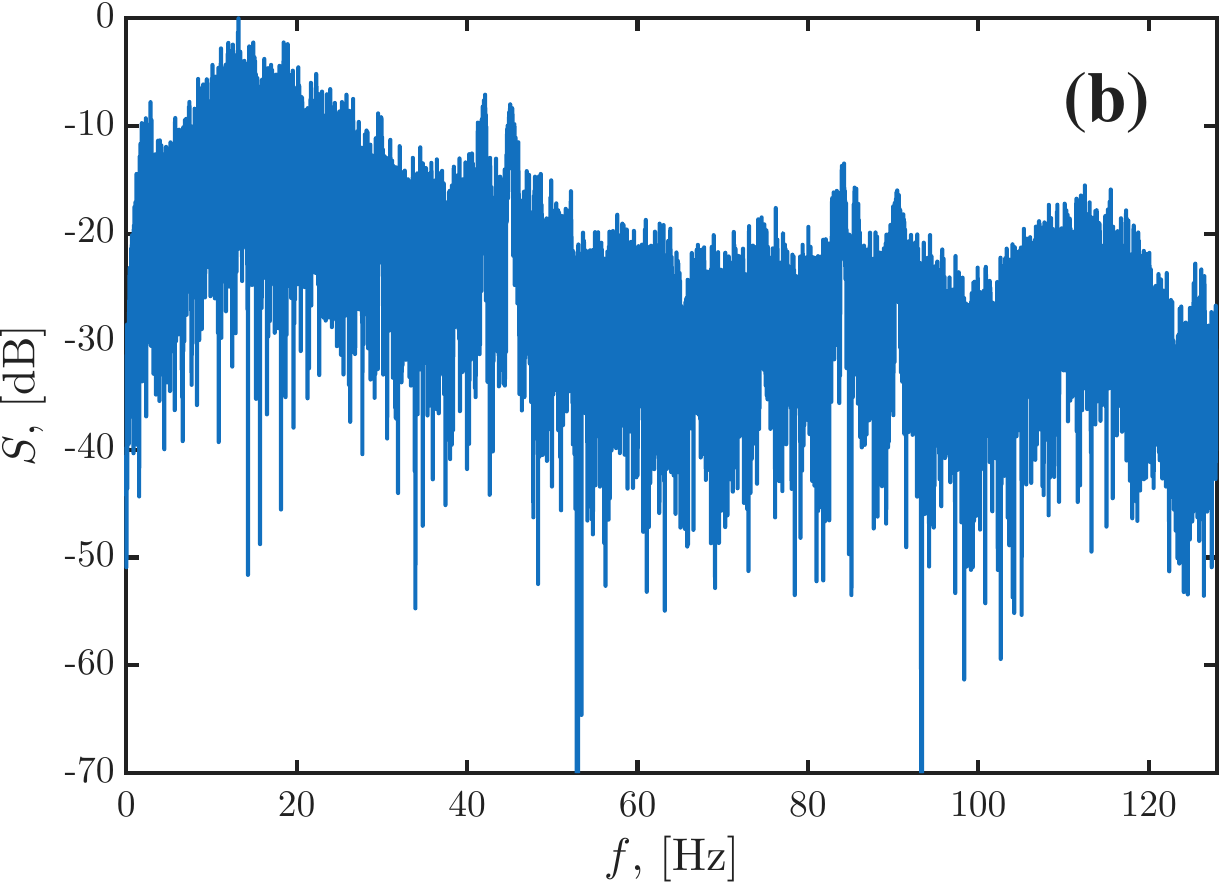}  	
\caption{(a) An example of the accelerometer output, a time trace $a(t)$, of a vehicle vibration. (b) The amplitude spectrum $S(f)$ of a single time trace $a(t)$ shown in Figure (a).  } 
\label{figure4}
\end{figure}
 
In contrast to the examples in the previous subsection, vibrations in a vehicle moving on a road are non-stationary and random due to the road surface. It means the vibration magnitude and its spectral content change in time. Additionally, the vehicle's manoeuvring and speed, as well as the vehicle's type, play a significant role. Figure \ref{figure4} shows an example of a vehicle vibration and its amplitude spectrum. An accelerometer (Siemens triaxial integrated circuit piezoelectric accelerometer) located near a wheel (the suspension's control arm) was used \cite{Shen:25} to obtain the acceleration signal shown in Fig.\,\ref{figure4}\,(a). Note that this accelerometer provides values in projection for each axis of 3-dimensional space, and the shown acceleration corresponds to the axis perpendicular to the road surface. The accelerations in the other directions have a comparable magnitude. The signal $a(t)$ (Fig.\,\ref{figure4}\,(a)) demonstrates an intermediate behaviour where small amplitude oscillations alternate with a spike and bursts of a larger amplitude. These high-amplitude oscillations are linked to the uneven road surface, and their magnitude can reach 20 g or more. The spectrum $S(f)$ (Fig.\,\ref{figure4}\,(b)) is broadband without identifiable periodic components. However, the spectrum is not completely flat. Two wide peaks are visible around 15 Hz and 115 Hz. These peaks reflect the vehicle construction features. The BaiQi Lite vehicle was used to collect the data. In this vehicle, the suspension control arm is relatively rigid, allowing the accelerometer to accurately reflect road-tyre interaction without significant distortion. 

\begin{figure}[!tb]
\centering
	\includegraphics[width=0.48\columnwidth]{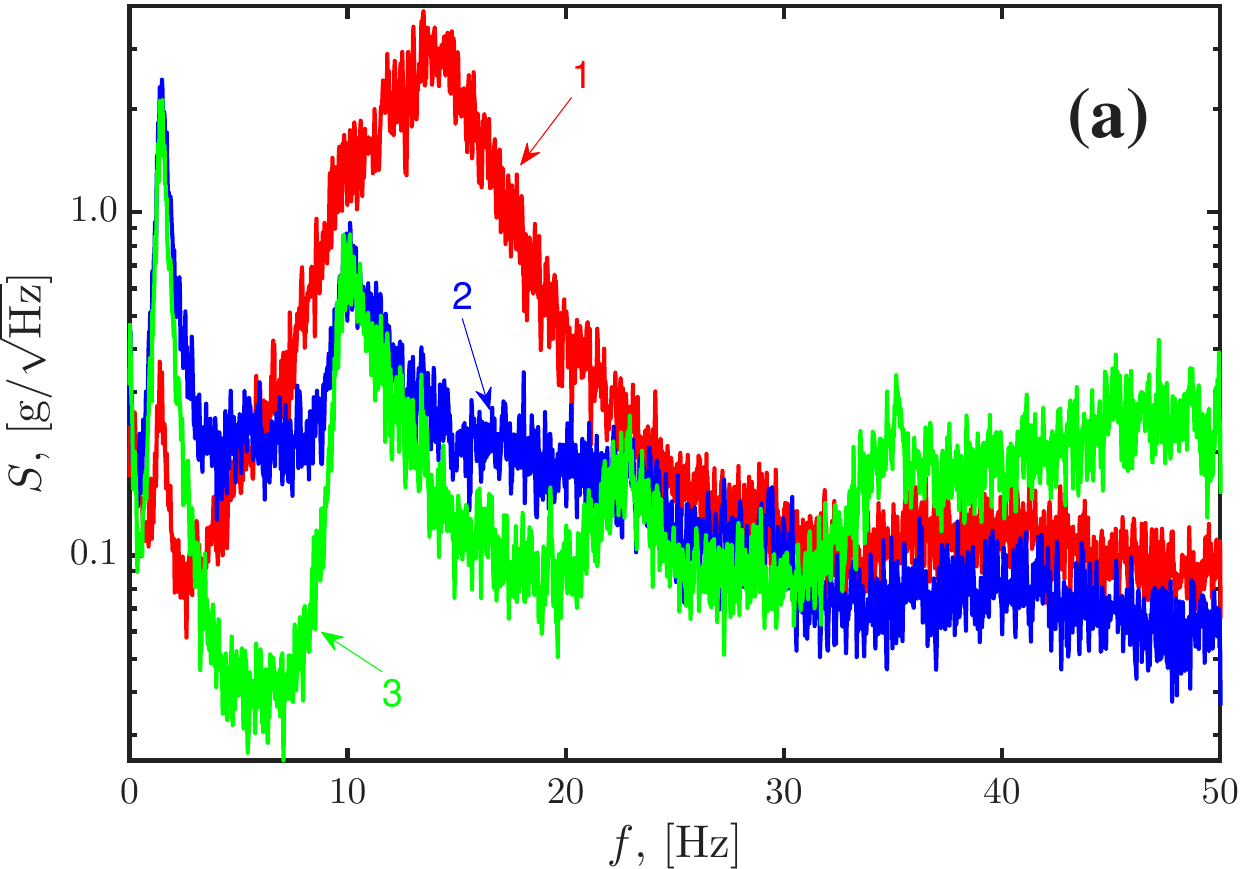} ~~
	\includegraphics[width=0.48\columnwidth]{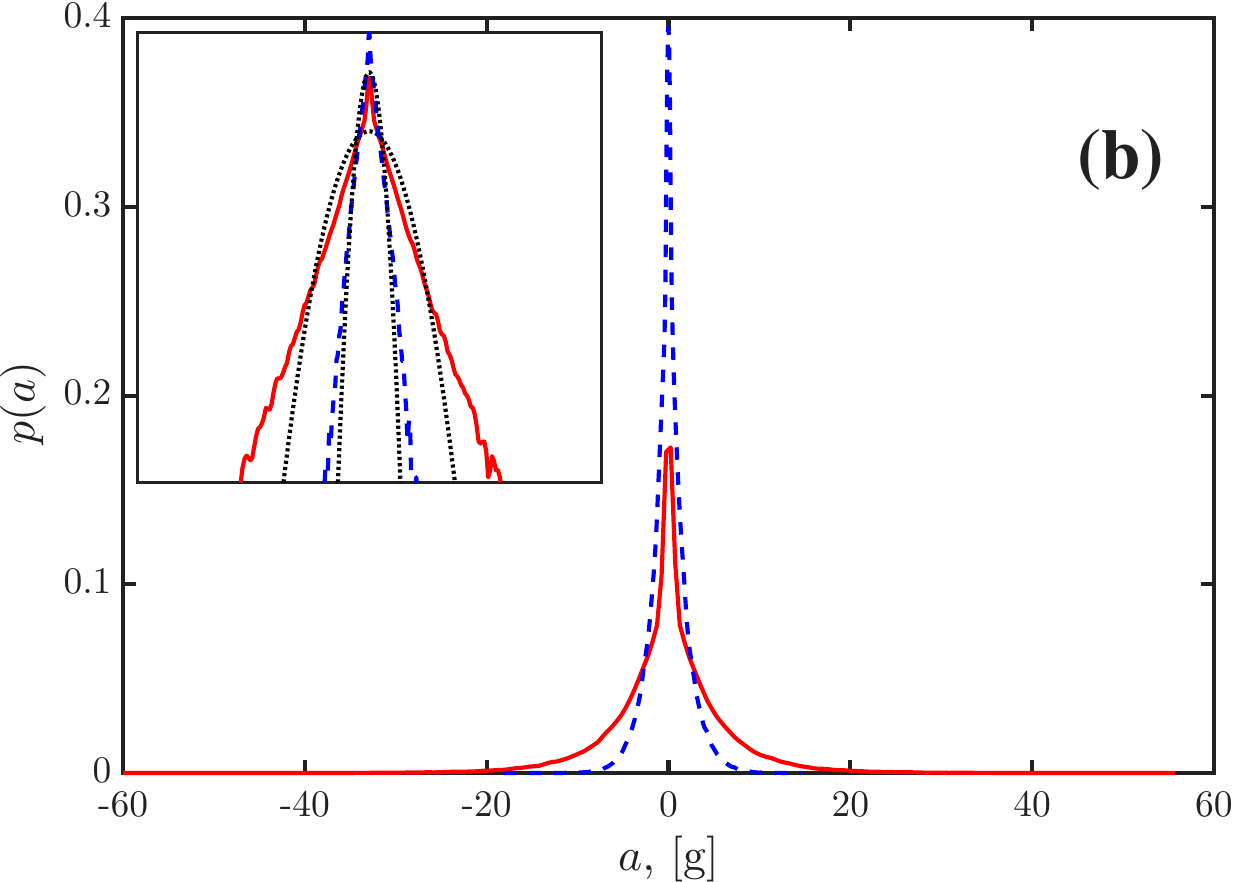}  		
\caption{(a) The amplitude spectrum $S(f)$ for acceleration time traces collected in different locations. Lines 1 (red), 2 (blue), and 3 (green) correspond to the suspension's control arm, the car body above the suspension, and the car's dashboard, respectively. 
 (b)  The probability density functions $p(a)$ for accelerations collected at the suspension's control arm (the solid red line) and the car body above the suspension (the dashed blue line).  The inset shows both functions $p(a)$ on a logarithmic y-axis. In the inset, the x-axis range is the same as in the main figure. The dotted black lines show the fitted normal probability density function. 
} \label{figure5}
\end{figure}

Now, let us consider another dataset \cite{car_data2}, obtained by driving a Volkswagen Saveiro along different roads with unpaved, cobblestone, and asphalt surfaces for 13.81 km over 24 minutes at $100 $ Hz sampling rate. The triaxial accelerometer (InvenSense MPU-9250) was used to collect vibration data from three locations: the suspension's control arm, the car body above the suspension, and the car's dashboard. The acceleration signal $a(t)$ in the perpendicular direction to the road surface is considered at each location. The amplitude spectrum for accelerations collected in different locations is shown in Figure~\ref{figure5}\,(a) using the logarithmic scale for the y-axis. To estimate the spectrum, the signal $a(t)$ was divided into short windows (4096 data points) and the corresponding power spectral density was averaged. The spectra confirm that the vibrations are random, with no regular (harmonic) components. However, depending on the accelerometer's location, the shapes differ. Vibrations of the control arm exhibit a dominant, wide but narrowband peak associated with the arm's cantilever motion. The peak occurs at 13 Hz. There is another peak at 1.5 Hz, but with significantly less magnitude. For two other locations, above the suspension and dashboard, the spectrum demonstrates two distinct narrowband peaks at 1.5 Hz and 10 Hz. The first peak reflects the suspension dynamics, and the second is linked to the combined tyre-suspension reaction to the road surface. 

The probability density function $p(a)$ for the acceleration signal $a(t)$ for two locations: the control arm and car body, is shown in Figure~\ref{figure5}\,(b). Note that $p(a)$ for the dashboard is nearly the same as for that for the car body, and, therefore, it is not shown. The probability function is unimodal and distinct from the Gaussian (normal) distribution at all locations. The range of the car body's acceleration is significantly narrower than that of the control arm. It means the suspension significantly reduced and damped the road-induced vibration. The tails of $p(a)$ are wider than those for the Gaussian one, which reflects the character of road action on the vehicle. As shown in Fig.\,\ref{figure4}\,(a), the road irregularities result in large amplitude perturbations of short durations. 
 
These two datasets of vehicle vibration show that broadband perturbations at the road surface transform into narrowband vibrations, with peaks that depend on the vehicle structure and the locations where vibrations are scavenged. The vibrations are random, with a distribution that is not normal.

 \subsection{ Summary of ambient vibration properties  }
 
The examples presented above of ambient vibration are far from providing a comprehensive picture of what the vibration could be. Nevertheless, they highlight the key properties of vibration and show how vibration can be described. In general, ambient vibration can be viewed as a random process composed of multiple components. The applied consideration aligns well with the generic approach outlined in the book \cite{bendat_random_2011} for analysing data. A combination of the spectrum and the distribution of data characterises two independent data properties, which can be utilised in developing ambient vibration models. The ambient vibration can include regular and random components, which appear as peaks and a background in the spectrum, respectively. The regular component can be a periodic signal $a(t)=\sum_{n=1}^{m} A_n \sin (n \Omega t+\varphi_n) $, with the fundamental frequency $\Omega$ or a combination of several independent periodic signals. Such a combination is called a quasiperiodic signal. For the regular component, we need to specify a set of frequencies and their corresponding amplitudes in the signal model. A random component requires specifying a shape in the spectrum and the probability density function. The considered examples show that the spectrum can be broadband, with the magnitude nearly the same in the entire frequency range (see Fig.\,\ref{figure4}\,(b)), or it can include narrow peaks with a peak's maximum located at a particular frequency. The probability density can be close to or very distinct from the normal one. The normal distribution is considered the preferred model, since many processes resulting from the action of weakly interacting factors are normally distributed. Also, the use of the normal distribution significantly simplifies modelling and provides a large volume of rigorous results for consideration. 
 
The examples also show that the same vibration source (road surface) can produce differently shaped vibrations depending on the vehicle and the location where ambient vibrations are collected. It means that each object produces a different vibration. Ideally, an energy harvester needs to be optimised for each vibration source. However, as will be apparent from the considerations below, the development of such an adaptive harvester is an open question. Therefore, the harvester design should focus on typical vibrations, which often means those with an appropriate mathematical description and those already used in engineering practice. The last conclusion also holds for a typical harvester, which is developed based on previous experience. 

 \section{ Cantilever as a typical harvester  }

\begin{figure}[!tb]
\centering
	\includegraphics[width=0.2\columnwidth]{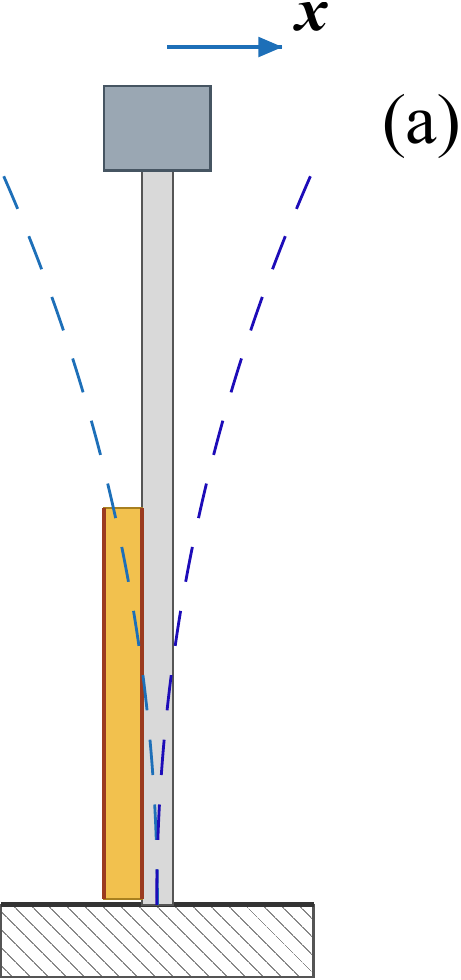} ~~
	\includegraphics[width=0.48\columnwidth]{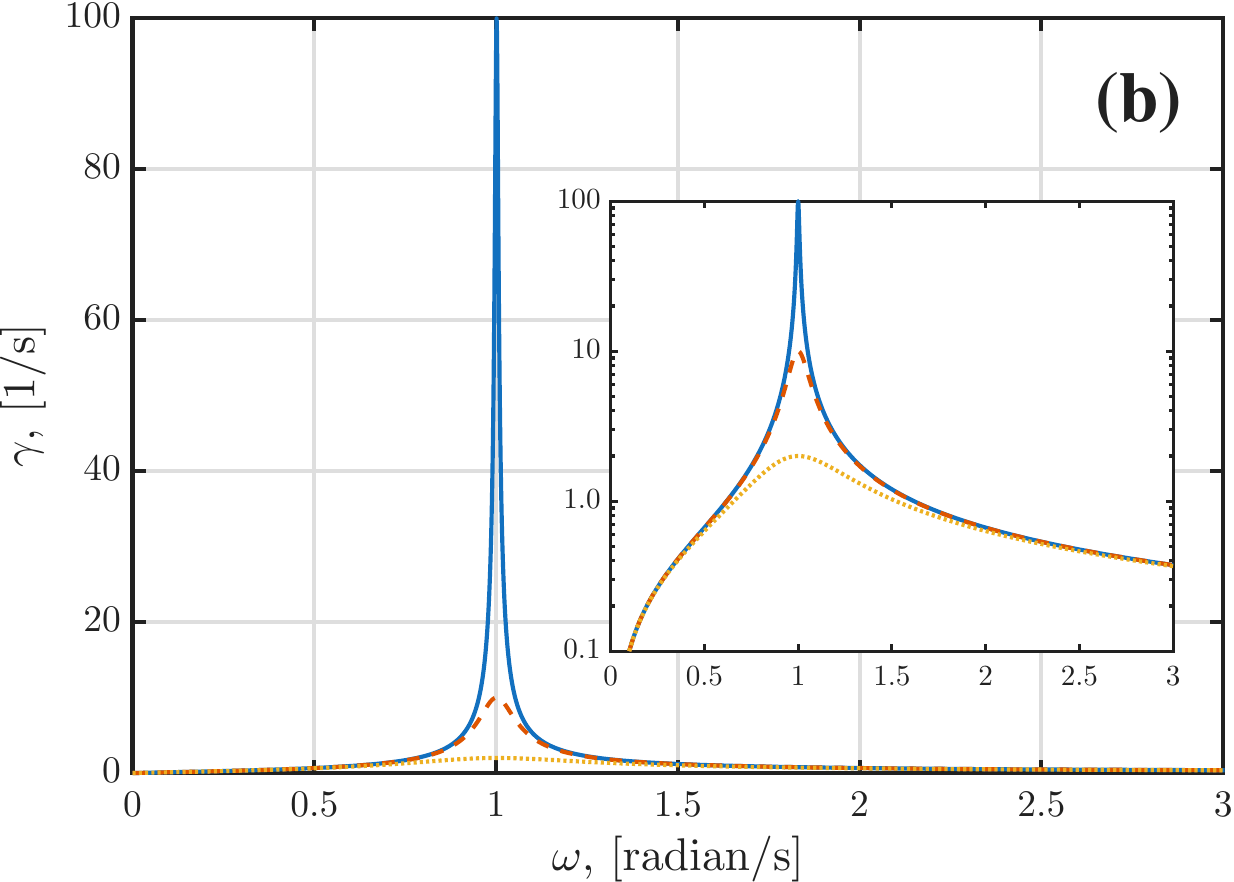}  
\caption{(a) Schematic picture of the piezoelectric harvester. The piezoelectric layer is shown attached to the left side of the cantilever beam. The dashed lines show beam configurations when the cantilever bends. (b) The gain $\gamma$, the amplitude frequency response, is shown for $\alpha=0.01$, $\alpha=0.1$ and $\alpha=0.5$ with the solid, dashed and dotted lines, respectively. The inset shows the frequency response on a logarithmic scale.} 
\label{figure6}
\end{figure}

The accelerometers used to characterise the vibration in the example above show that vibrating objects produce a force acting over a small area, and a smaller attached object (an accelerometer) has diminishing back-action on the vibration. It means if a harvester is smaller and lighter than the vibrating object, the object's acceleration transforms to the force, which is proportional to the harvester mass: $F(t) \propto m a(t)$, and the force is applied at a point. Here, we arrive at the question of what the harvester should be to utilise this force. The answer is based on previous engineering solutions that are simple, durable, well-understood, and tunable. Therefore, a cantilever beam, which is a beam fixed at one end, free at the other and having a tip mass, is the choice number one for an energy harvester. The fixed end, the base, is attached to the vibrating object and the force $F(t)$ applied to the base leads to the cantilever oscillations. The beam properties, such as geometry, materials, and the tip mass, are adjustable design parameters. During cantilever bending, the strain has opposite signs on the beam's opposite faces: tension on one face and compression on the other over each half-oscillation. Such a strain distribution is ideal for a piezoelectric layer bonded to the beam face, since it yields a strong, coherent voltage output rather than strain cancelling out across different regions, as it does in more complex shapes, such as a plate. Additionally, compared to other mechanical shapes, a cantilever motion is nearly one-dimensional, and it gives the largest deflection and, therefore, the largest strain for a given base acceleration. The larger strain results in greater electrical power from a piezoelectric layer. Thus, a cantilever with an attached piezoelectric layer is a typical energy harvester. A schematic picture of the piezoelectric harvester is shown in Fig.\,\ref{figure6}\,(a). 

Cantilever's dynamics are well described by an oscillator model with a single degree of freedom:
\begin{align}
\label{eq_cant}
	m \ddot{x} + F_d( \dot{x}) +F_s  (x ) = - F(t)      \ , 
\end{align}
where $x$ represents the beam motion, it can be a displacement of the tip or an angle describing the beam bending, $m$ is an effective cantilever mass that includes the beam and tip, $F_d$ is a damping force resulting mainly from energy dissipation within the beam's material, $F_s$ is a stiffness force produced by the beam as a reaction to bending from an equilibrium state, and $F(t)$ is the ambient vibration. Typically, $m$, $F_d$, and $F_s$ are identified from measurements using well-established techniques \cite{Kerschen:2006}, and they vary with the cantilever dimensions and the beam's material. In the piezoelectric harvester, the piezoelectric layer (an electrical part) is coupled with the cantilever (a mechanical part) to form the following model:
\begin{align}
	\begin{split}
&	m \ddot{x} + F_d( \dot{x}) +F_s  (x ) +\theta_p V = - F(t)      \ ,  \\
&		 C_p \dot{V}+ \frac{V}{R} = - \theta_c \dot{x} \  .
	\end{split}
	\label{eq_eh}
\end{align}
Here $V$ is the output voltage of the piezoelectric layer and $C_p$ is its capacitance,  $\theta_c$ and $\theta_p$ are the electromechanical coupling coefficients, and  $R$ is the load resistance representing a power-management electrical circuit. The coefficients $\theta_c$ and $\theta_p$ and capacitance $C_p$ are defined from measurements. The coefficient $\theta_p$ is negligible for a large cantilever having dimensions of centimetres, and its influence is growing (but still it is limited) in the micrometre range.  For the selected piezoelectric layer, the amplitude of the cantilever oscillation is the primary factor determining the output electrical power. Therefore, the damping, $F_d$, and stiffness, $F_s$, forces need to be adjusted to maximise the harvester efficiency, and these forces are the main factors in the cantilever design.

\subsection{Linear resonance}

In engineering, a linear design and corresponding linear models are dominant~\cite{Thorby:2008}. There are many reasons for such a situation. For example, linear systems can be comprehensively analysed using analytical solutions; there are many tools for obtaining and utilising these solutions. So, the most widely used harvester model assumes the linear form for the forces: $F_d=\beta \dot{x}$ and $F_s= k x$, which leads to the linear harvester:
\begin{align}
	\begin{split}
&	m \ddot{x} + \beta  \dot{x} + k x +\theta_p V = - F(t)      \ ,  \\
&		 C_p \dot{V}+ \frac{V}{R} = - \theta_c \dot{x} \  .
	\end{split}
	\label{eq_lineh}
\end{align}
Assuming that the back action of the piezoelectric layer of the cantilever is small, the consideration can be focused on the harvester mechanical part, which is described by the linear oscillator:
\begin{align}
\label{eq_linosc}
	\ddot{x} + \alpha  \dot{x}  + \omega_n^2  x = - a (t)      \ , 
\end{align}
where the forces are normalised by $m$, and $\alpha=\frac{\beta}{ m}$ is the normalised damping coefficient, $\omega_n=\sqrt{\frac{k}{m}}$ is the oscillator natural frequency, and $a(t)$ is the ambient vibration acceleration if we assume $F(t)= m a(t)$. The response of the linear oscillator (\ref{eq_lineh}) to any form of $a(t)$ can be obtained by using the transfer function \cite{Inman:2014}, and it can be understood by considering the oscillator's frequency response. Analytically, the frequency response can be found by considering the external input and the oscillator displacement in the complex form: $a(t)= A \exp (i \omega t)$ and $x(t)= X \exp (i \omega t)$, respectively. Note that $X=|X|\exp(i \varphi)$ is a complex variable. The coefficient $i=\sqrt{-1}$ is the imaginary unit.  The harvester voltage output is proportional to the amplitude of velocity (see the second equation in (\ref{eq_lineh})), therefore the frequency response of the velocity  $\dot{x}(t)= i \omega X \exp (i \omega t)$ defines the harvester output.  The frequency response is a complex function, and the gain is its modulus, which defines the ratio of the output to input amplitudes, over the entire frequency range, formally, for $\omega \in [0, \infty)$.  The expression for the gain is the following:
\begin{align}
\label{FR_gain}
	\gamma (\omega) =\left| \frac{i \omega X}{A} \right| =\frac{\omega }{\sqrt{ \left( \omega_n^2 -\omega^2 \right)^2 + \alpha^2 \omega^2 } }      \  , 
\end{align}
and it defines the amplitude frequency response of the oscillator (\ref{eq_linosc}). The expression (\ref{FR_gain}) describes linear resonance, in which the response is maximal at a particular frequency linked to the oscillator's parameters and independent of the input amplitude. For cantilever velocity $\dot{x}$,  the maximal gain is observed on the resonance frequency $\omega_m=\omega_n=\sqrt{\frac{k}{m}}$, which is equal to the natural frequency and does not depend on the damping coefficient $\alpha$. However, $\alpha$ determines the magnitude of the maximal gain
\begin{align}
\label{FR_gain_max}
	\gamma_m=\gamma(\omega_m)=  \frac{1}{\alpha}    \ . 
\end{align}
The gain $\gamma$ shows how efficiently the vibration acceleration $a(t)$ is transformed into the cantilever velocity $\dot{x}(t)$, and it has the dimension of $1/\text{s}$.  The amplitude frequency response (gain) for several values of $\alpha$ is shown in Fig.\,\ref{figure6}\,(b) and demonstrates two key features of the linear harvester. First, the linear harvester is efficient in a narrow frequency range around the cantilever's natural frequency. Second, the smaller the damping, the more efficient the harvester. So, if the ambient vibration is close to the single harmonic, a dominant peak at a particular frequency $f_p$ in the spectrum shown in Fig.\,\ref{figure2}\,(b), the conducted analysis identifies design parameter $\omega_n$, which should be tuned to match $2 \pi f_p$, and parameter $\alpha$, which should be maximally reduced. 

Note that the gain is an experimentally measured quantity. In the experiment, the input is a harmonic signal $a(t)=A \sin(\omega t)$, produced, for example, by a vibrational shaker. Sweeping the frequency over a wide range and measuring the steady-state amplitude of the velocity oscillation yields the gain. Alternatively, the input can be white noise, a broadband noise with constant spectral magnitude at each frequency. Then, collecting many velocity trajectories and estimating an average spectrum using a similar procedure as for Fig.\,\ref{figure2}\,(b), the gain is evaluated. 
So the modelling consideration provides consistent guidance with verified steps for practical implementation.  The engineering task consists of selecting appropriate beam dimensions, material, and tip mass; manufacturing and testing; and then amending the harvester to meet the design criteria. While the analytical solution to the design problem is relatively straightforward, implementing it in practice can be challenging. For example, the considered vibration examples illustrate that ambient vibration can include several independent harmonic signals, contain noise, and is often purely random. In the case of several harmonic signals, the strategy is to use a combination of linear harvesters \cite{Beeby:11}. For broadband random vibration, the linear harvester is not an optimal solution.

The main practical issue is that the cantilever exhibits a linear response over a limited range of beam displacement, and the response becomes highly nonlinear as the cantilever dimension is reduced \cite{Lifshitz:2008}. Therefore, nonlinear damping and stiffness forces must be included in the design considerations.  In contrast to linear systems, nonlinear systems are rarely analytically tractable and must be simulated numerically. Additionally, there is no universal description of nonlinearity in the sense that each nonlinear function is unique. Therefore, in engineering, some types of nonlinearity are considered, and these types are associated with particular properties of the considered system. In the harvester model (\ref{eq_eh}), the damping force is typically considered as linear. In many cases, it's a simplification that allows us to focus on the nonlinearity in the stiffness force $F_s (x)$. Note that experiments show that the nonlinearity in stiffness has a greater effect than the damping nonlinearity in the harvester. 

\subsection{Nonlinear resonance}

\begin{figure}[!tb]
\centering
	\includegraphics[width=0.48\columnwidth]{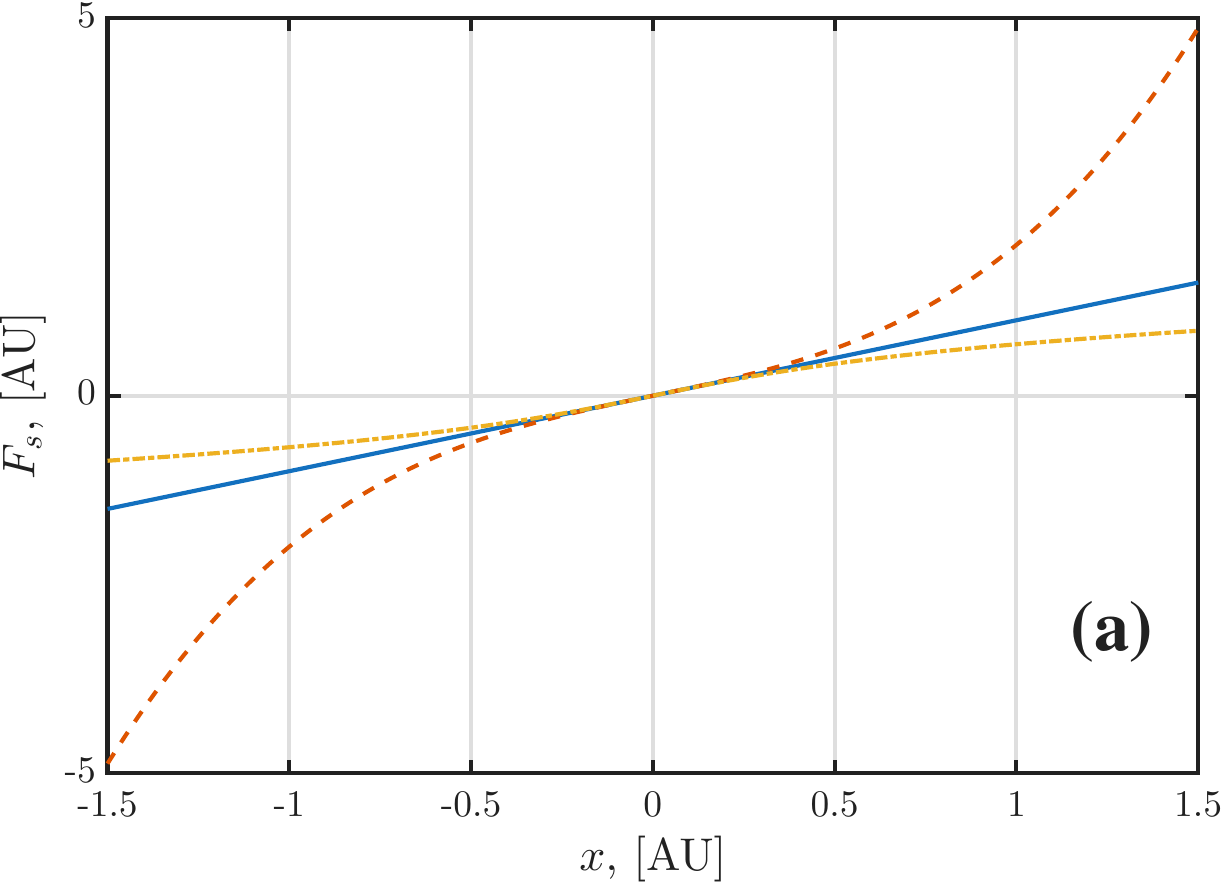} ~~
	\includegraphics[width=0.48\columnwidth]{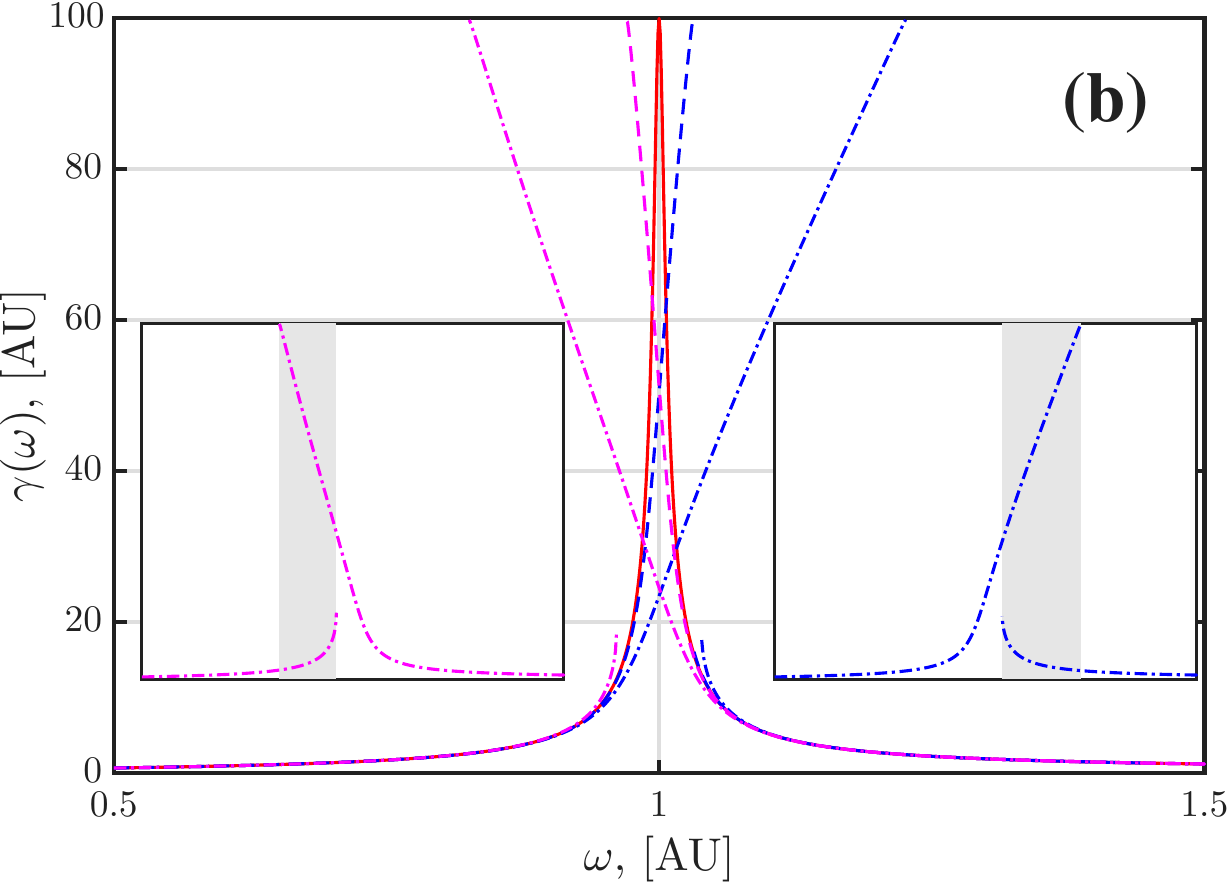}  
\caption{ (a) The cantilever's stiffness force is shown for the linear case (the solid line), and for hard (the dashed line) and soft (the dashed-dotted line) types of the nonlinearity. (b) The amplitude frequency response is shown for the hard (curves bend to the right, blue) and soft (curves bend to the left, magenta) types of nonlinearity. The dependences $\gamma(\omega)$ are shown for different input amplitudes: $A=0.0001$ (the solid line), $A=0.003$ (the dashed line) and $A=0.01$ (the dashed-dotted line).  The insets show $\gamma(\omega)$ for $A=0.01$ with the shadowed regions corresponding to the hysteresis loop.  } \label{figure7}
\end{figure}

In engineering, the nonlinear stiffness is classified into two types: hard and soft nonlinearity. The differences between these types are illustrated in Fig.\,\ref{figure7}\,(a). The linear force $F_{s,l}= k x$ assumes that the effort to move the system by the distance $x$ from the initial state, $x=0$, is proportional to the distance. Note that typically the linear representation $F_{s,l} $ of the arbitrary force $F_s(x)$ is obtained by the linearisation of $F_s (x)$ around a state, $x=0$, in the considered case (the solid line in Fig.\,\ref{figure7}\,(a)). If the actual (not linearised) force is larger than the linearised prediction (the dashed line in Fig.\,\ref{figure7}\,(a)), it is actually harder to move; the corresponding nonlinearity is called {\it hard} nonlinearity. In the opposite situation, when the force is smaller (the dashed-dotted line in Fig.\,\ref{figure7}\,(a)), it is a {\it soft} nonlinearity. The hard type can be observed across the entire range of displacements and is a typical mechanical system nonlinearity. The soft type is observed over a limited range of displacements, with either some hardening regions around or regions with negative stiffness.

Let us consider how the nonlinearity affects the cantilever's frequency response. More precisely, we consider a qualitative change, a tendency in the changes. For a quantitative analysis, we have to consider a particular actual cantilever, and the outcome will not be generic. Also, we assume that the damping force is linear. The nonlinear cantilever model has the following form:
\begin{align}
\label{eq_nonlinosc}
	\ddot{x} + \alpha  \dot{x}  + f_s ( x ) = - a (t)      \ . 
\end{align}
The hard type has a canonical cubic form: $f_{s,h}(x)=x+\mu x^3$, $\mu>0$. This form is associated with the Duffing oscillator \cite{Kovacic:2011}, the first model used to describe nonlinear vibration. In contrast, there is no widely used form for the soft type.  Here, for the soft type, we consider the inverse cubic form: $f_{s,s}(x)=y(x)$, where $y$ is a root of the cubic equation $y+\mu y^3-x=0$ , $\mu>0$. The root is unique and defined by modified Cardano's formula:
 \begin{align}
\label{Cardono_formula}
	y=C-\frac{1}{3 C \mu}, \ \ \ C= \sqrt[3]{-\frac{x}{2\mu} + \sqrt{ \frac{x^2}{4\mu ^2} + \frac{1}{27\mu^3} } }      \ . 
\end{align}
To estimate the amplitude frequency response, $\gamma(\omega)$, the input has the form $a(t)=A \sin(\omega t)$ and the differential equation is solved numerically using the Runge-Kutta 4th-order method \cite{Butcher:16} with the constant step size. 
The estimation mimics the experimental setup by performing a frequency sweep over the range $\omega\in[\omega_1, \omega_2]$ with a frequency step $h_{\omega}$. For the first frequency value $\omega=\omega_1$, the initial conditions are selected at an equilibrium state $(x=0, \dot{x}=0)$, and the system (\ref{eq_nonlinosc}) is simulated for an initial transient time interval of 500 periods of the input force, and then for a further 100 periods assuming the system arrived at the steady state. The gain is defined by the following expression:
\begin{align}
\label{gain_num}
	\gamma (\omega )=\frac{1}{A} \sqrt{ \frac{2}{\tau} \int_{t=0}^{t=\tau}  \dot{x}^2 dt }      \  ,
\end{align} 
where $\tau=100 \frac{2\pi}{\omega}$. This integral is replaced with a summation in a computer code. Then the frequency is increased by the step $h_{\omega}$, and the values of $x$ and $\dot{x}$ at the last simulation time instance are used as initial conditions. After the frequency reaches the interval end $\omega=\omega_2$, the sign of $h_{\omega}$ is reversed, and the estimation is performed for frequencies reduced from $\omega_2$ to $\omega_1$.  In this manner, the input frequency initially sweeps from small to larger values, then reverses direction. Frequency sweeping in the reverse direction is irrelevant for the linear cantilever and allows us to check for multiple states in the nonlinear case. 

The amplitude frequency responses for the cantilever with hard and soft types of nonlinearity are shown in Figure~\ref{figure7}\,(b). Let us stress that the frequency response of the linear system (\ref{eq_linosc}) does not depend on the input amplitude $A$. For small values of $A$, when a linearised representation $f_s(x)=k x$, $k=1$, of the stiffness forces $f_{s,h} (x)$ and $f_{s,s} (x)$ is valid, the frequency responses of the nonlinear systems are the same as for the linear one. The increase in $A$ leads to a nonlinear response distinct from the linear resonance curve. Depending on the type of nonlinearity, the frequency response curves bend to the left for soft nonlinearity and to the right for hard nonlinearity. The curves become asymmetric and wider. The location of the resonance frequency, where the gain is maximal, shifts to the left or right depending on the type of nonlinearity as the input amplitude $A$ increases. For the hard type, the resonance occurs at higher frequencies than in the linear case, whereas for the soft type, it occurs at lower frequencies. Starting with some magnitudes of $A$, the frequency curves demonstrate discontinuities where the gain $\gamma$ jumps between values and two partially overlapped branches with a discontinuity are formed. One branch is observed when the frequency sweeps from small to large values, and another when sweeping is performed in the opposite direction. The area between these two discontinuities is called a hysteresis loop (shadow areas in the insets in Fig.\,\ref{figure7}\,(b)), where two different gains can be observed depending on the initial conditions. The described nonlinear response is known as the nonlinear resonance and, if the hysteresis is present, it is often called the jump resonance (see the book \cite{Nayfeh:1981} for more details). Note that for a large amplitude, one can observe a more complex response, including dynamical chaos~\cite{Moon:2004}. 

The nonlinear response has a profound effect on the design of the energy harvester. In the linear case, it was sufficient to match the harvester's natural frequency to the vibration frequency to maximise harvester efficiency. If nonlinearity plays a role, the vibration magnitude must be accounted for in the design. The latter means that a comprehensive understanding of the cantilever's nonlinearity as a function of its parameters, such as beam material, dimensions, and tip mass, is required. Further, the design process must include a robust, validated harvester model to predict the nonlinear response. Note that while all tools for design, including nonlinear predictive modelling, are available and explored, the routine application of nonlinear design is scarce, as linear design remains the main engineering paradigm. 

\section{ Interplay between ambient vibrations features and harvester nonlinearity   }

In the subsection above, the nonlinearity acts destructively, shifting the cantilever's resonance frequency and thereby reducing the output amplitude. Simultaneously, many researchers view nonlinearity as an opportunity to overcome the narrowband limitation of linear harvesters and to explore nonlinear designs for broadband vibration. Reviews \cite{Jia:2020} reflect recent developments.  However, the random component of ambient vibration is often overlooked, with the main focus on periodic inputs or combinations of periodic inputs. Including noise in the consideration can dramatically alter the predictions of noise-free modelling. Also, in many cases, the noise is the only driving force of ambient vibration.  In the subsections below, we consider several examples of how the interplay between noise and the cantilever nonlinearity affects the harvester performance. 

\subsection{ Harmonic oscillations with varying frequency }

While nonlinear resonance shifts the resonance frequency, it widens the frequency range over which the gain is large. Therefore, some publications (see \cite{Jia:2020} and the references therein) suggest utilising the broadband nonlinear response to improve the harvester performance in a situation where the input frequency varies in some frequency range. Let us consider the harvester model (\ref{eq_eh}) in the normalised and dimensionless form:
\begin{align}
	\begin{split}
&	\ddot{x} + \alpha  \dot{x} + f_s (x) +\psi  v = - a(t)      \ ,  \\
&		  \dot{v}+ \lambda v = - \kappa \dot{x} \ ,
	\end{split}
	\label{eq_neh}
\end{align}
where $\alpha=\beta/(m \omega_n^2)$, $f_s(x)=F_s(x)/(m \omega_n^2)$, $\psi=\theta_p/(m \omega_n^2)$, $\lambda=1/(RC \omega_n^2)$, $\kappa=\theta_c/(C \omega_n^2)$ and $a(t)=F(t)/(m \omega_n^2)$. The natural frequency $\omega_n$ is determined for the linearised model. Using $\omega_n^2$ normalisation, we make a dimensionless natural frequency of the harvester close to 1, and all variables are presented in arbitrary units [AU], which is typical in modelling.  The following parameter values adapted from literature \cite{Erturk:09,Litak:10,Khovanova:2014} are used: $\alpha=0.01$, $\psi=0.05$, $\lambda=0.05$, and $\kappa=0.5$.  

\begin{figure}[!tb]
\centering
	\includegraphics[height=0.31\columnwidth]{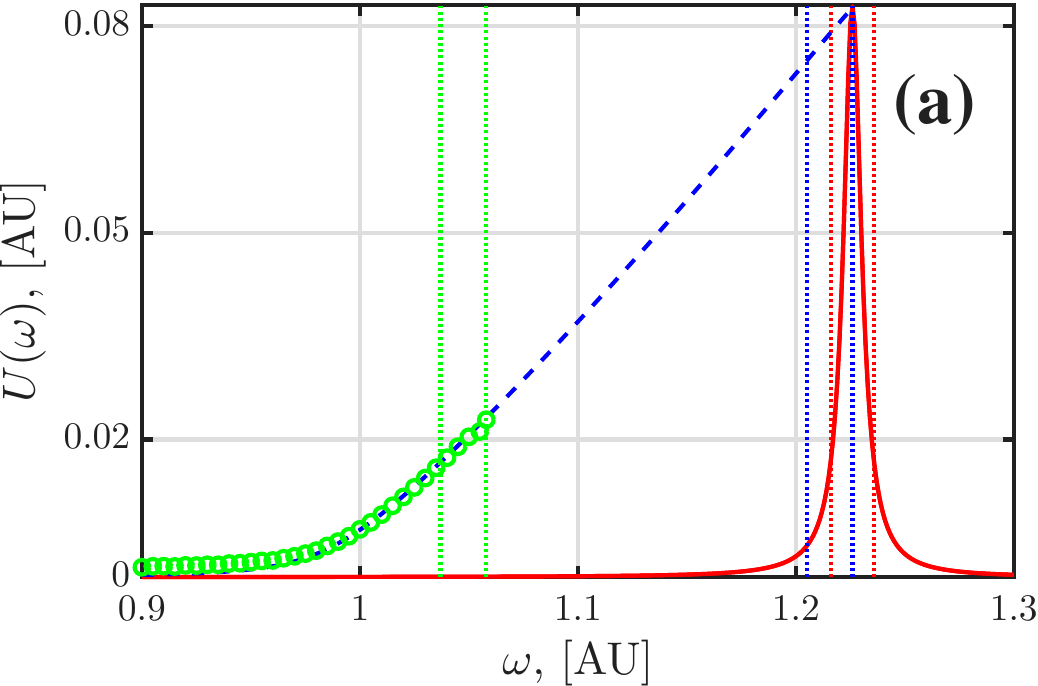} ~~
	\includegraphics[height=0.31\columnwidth]{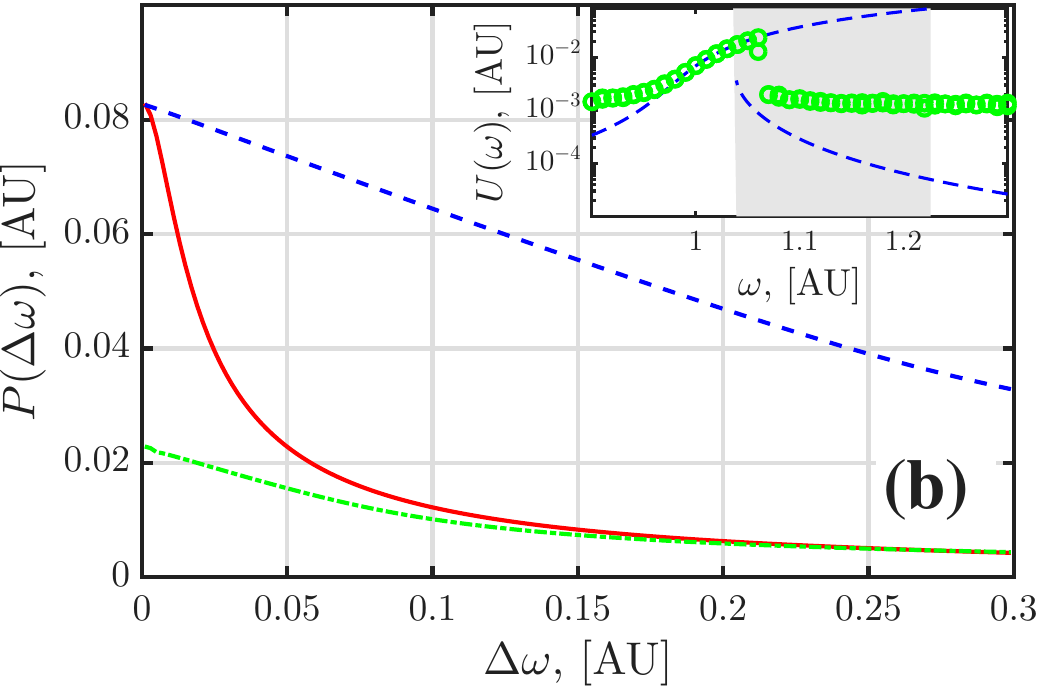}  
\caption{ (a) The harvester output amplitude $U$ as a function of the input frequency $\omega$.  The solid red and dashed blue lines show $U(\omega)$ for the linear and nonlinear cantilever, respectively. The green marker $\circ$ shows $U(\omega)$ for a nonlinear cantilever in the presence of white noise. The vertical dotted lines denote the frequency range $\Delta \omega$ within which the input frequency $\omega$ varies. (b) The harvester output power $P(\Delta \omega)$ for time-varied input frequency $\omega$ within the range $\Delta \omega$ is shown for the linear harvester (the solid red line), the nonlinear harvester when the noise is absent (the dashed blue line) and present (the dashed-dotted green line). The inset shows the output amplitude $U(\omega)$ as a function of frequency. The dashed blue line and marker $\circ$ (green) show $U(\omega)$ for the nonlinear harvester in the absence and presence of noise, respectively. The shadow area denotes the hysteresis loop observed in the noise-free case. } \label{figure8}
\end{figure}

Let us assume that the input acceleration is a single harmonic $a(t)=A \sin(\omega t) $ and the frequency $\omega$ uniformly slowly varies in a frequency range $\Delta \omega\in[\omega_1,\omega_2]$, so the harvester output power $P$ is defined by the following expression:
\begin{align}
P (\Delta \omega) =\frac{1}{\Delta \omega} \int_{\omega_1}^{\omega_2} U^2 (\omega) d \omega \ ,
	\label{eh_pow1}
\end{align}
where the output amplitude $U (\omega)$ of the harvester for $a(t)=A \sin(\omega t) $ with not-varied $\omega$. Values of $U (\omega)$ are estimated in a similar way as the amplitude in the frequency response for the cantilever:
\begin{align}
\label{pw_num}
	U = \sqrt{ \frac{1}{\tau} \int_{t=0}^{t=\tau}  \dot{v}^2 dt }      \  .
\end{align} 
Further, consider the linear $f_s= k_1 x$ and nonlinear stiffnesses of the hard type $f_s= x+\mu x^3$, and select the value of $A=0.01$ such that the frequency response of the nonlinear harvester demonstrates the hysteresis. By setting $k_1=1.226$, the resonance frequency of the linear and nonlinear harvesters is the same (Fig.\,\ref{figure8}\,(a)).  The linear harvester performance is optimal if the middle of the range $\Delta \omega$ is located at the resonance frequency $\omega_m$.  For the nonlinear harvester, the range should be $\omega \in [\omega_m-\Delta \omega, \omega_m]$ for the most efficient performance. The selection of the frequency range is illustrated by vertical dotted lines in Fig.\,\ref{figure8}\,(a). The dependences $P(\Delta \omega)$ estimated using expression (\ref{eh_pow1}) confirm that the suggested advantage of the nonlinear design is valid and the nonlinear harvester has a larger power output for the single harmonic input with a time-varied frequency (Fig.\,\ref{figure8}(b)). For the non-varied frequency, i.e. $\Delta \omega=0$, the power is the same for both harvesters. As the width of $\Delta \omega$ increases, power $P(\Delta \omega)$ decreases sharply as $P(\Delta \omega) \propto (\Delta \omega)^{-2}$ for the linear system and more slowly $ P(\Delta \omega) \propto -\Delta \omega $ in the nonlinear case. Thus, the nonlinear harvester looks significantly more efficient than the linear one for the harmonic signal with a time-varying frequency. 

However, typical ambient vibrations include a strong random component. Let us repeat the same analysis with the broadband noise included in $a(t)$. White noise $n(t)=\sqrt{D}\xi(t)$ of intensity $D=A^2$ is added to the input.  So,  $a(t)=A \sin(\omega t) + n(t) $.  The white noise has a flat spectrum across the entire frequency range, that usually expressed via the correlation function: $\langle \xi(t) \xi(0)\rangle =\delta(t) $, where $\delta(t)$ is the delta function and the notation $\langle . \rangle$ means time averaging. Also, white noise has zero mean, $\langle \xi(t) \rangle = 0$, and is normally distributed. Since the noise source is present, the expressions (\ref{eq_neh}) are stochastic differential equations that require corresponding numerical methods \cite{Mannella:Heun}. The Heun method, a stochastic equivalent of the 2nd-order Runge-Kutta method, was used.

The selected noise intensity means that the noise is considerably weaker than the harmonic signal. The additional random component barely affects the linear harvester, and the dependence $P(\Delta \omega)$ is practically the same and therefore skipped. In contrast, the performance of the nonlinear harvester significantly degrades (see the dashed-dotted green line in Fig.\,\ref{figure8}\,(b)).   
The reason for such a dramatic change is the following. In the nonlinear resonance, the hysteresis loop (a shadow area in the inset of Fig.\,\ref{figure8}\,(b)) implies bistability, the coexistence of two stable oscillatory states separated by an unstable state. The stable states are reflected in the frequency response, and the unstable state is unseen. One of these stable states corresponds to large-amplitude oscillations (large $\gamma$), and another to small-amplitude oscillations (small $\gamma$). In the noise-free estimation of $P(\Delta \omega)$, the large amplitude state was used. The random force perturbs each state, and the most stable is observed. For the selected hard type of the nonlinearity, which is the most typical one in mechanical systems, the unstable state is close to the large-amplitude state, and a small-amplitude state is more stable~\cite{Khovanova:12}. So, as it is shown in the inset in Fig.\,\ref{figure8}\,(b), the large-amplitude state is not observed in a significant part of the hysteresis loop when noise is present. As a result, the advantage of the nonlinear harvester observed in the noise-free situation disappears in more realistic settings.  
Note that in the frequency range, close to the maximal power (see the inset in Fig.\,\ref{figure8}\,(b)), the coexisting states demonstrate similar stability, and noise induces switching between the states and the output amplitude $U(\omega)$ reduces. 

\subsection{ Bistable harvester  }

\begin{figure}[!tb]
\centering
	\includegraphics[height=0.32\columnwidth]{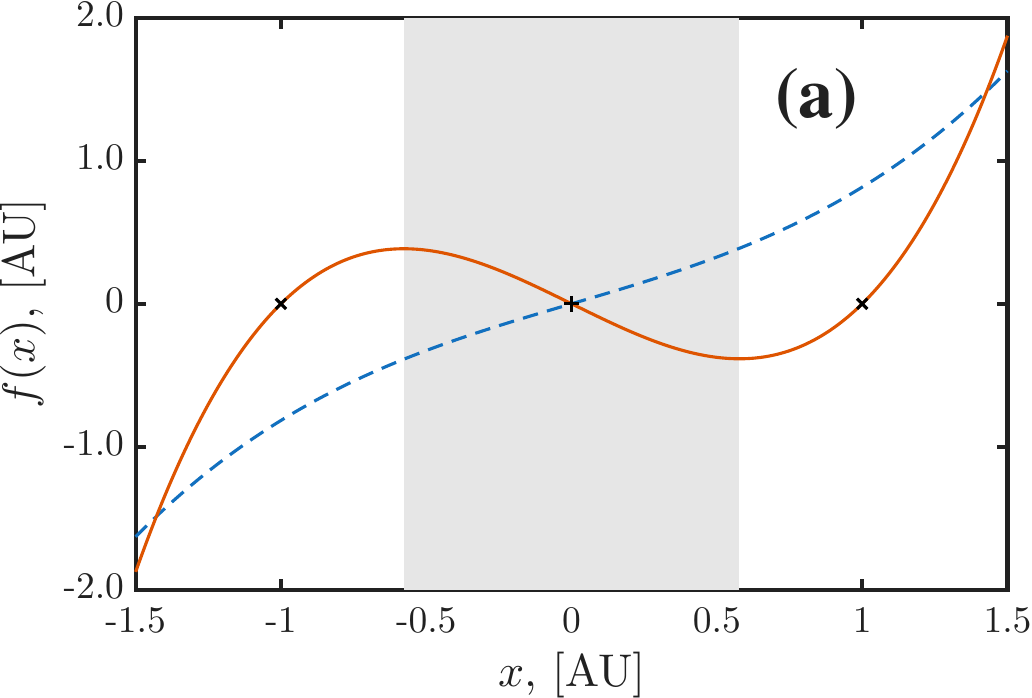} ~~
	\includegraphics[height=0.32\columnwidth]{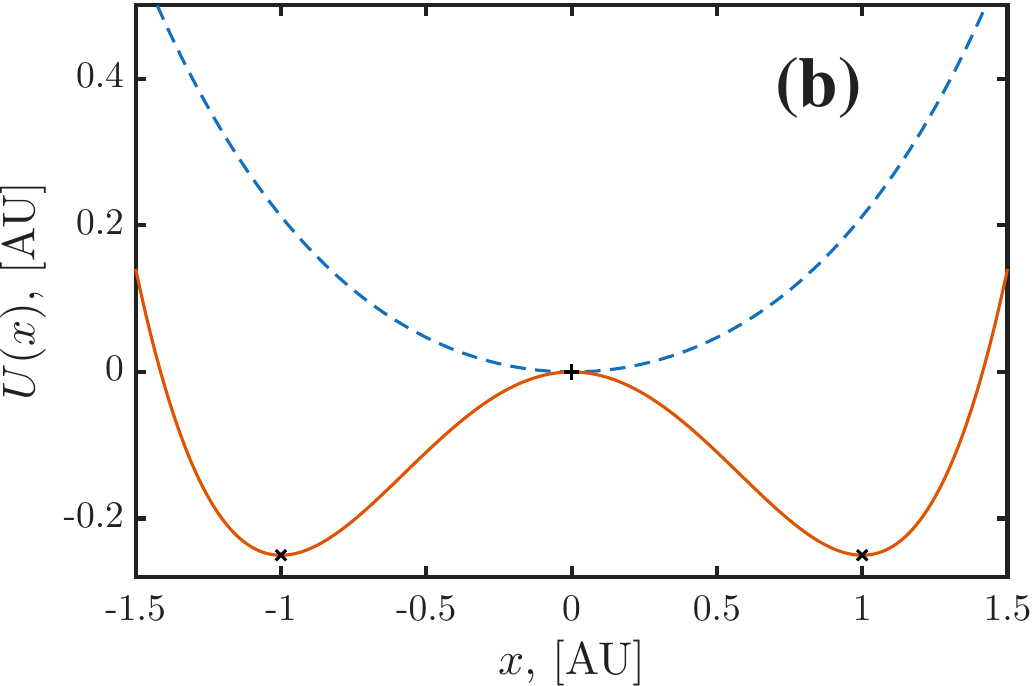}  
\caption{ The cantilever stiffness force $f(x)$ (Figure (a)) and the corresponding potential $U(x)$ (Figure (b)) for the monostable (the dashed line) and bistable (the solid line) cases. The markers $\times$ and $+$ denote the stable and unstable equilibrium states of the bistable harvester, respectively. The shadow area in Figure (a) denotes the region with the negative stiffness. } \label{figure9}
\end{figure}

In the considerations above, bistability in a nonlinear oscillator appears in the response to external harmonic forces, whereas the oscillator has a single state when the force is absent or small. Also, we see that bistability is not beneficial for energy harvesting. However, this observation does not imply that bistability is always a limiting factor, and the performance of bistable harvesters has been explored in many studies (see \cite{Jia:2020,Cottone:09} and references therein). The cantilever stiffness force $f(x)$ (the dashed line in Fig.\,\ref{figure9}\,(a)) is modified, for example, by introducing magnets that attract the tip, thereby creating two equilibrium states (the solid line in Fig.\,\ref{figure9}\,(a)). Often, in modelling, the following cubic function $f(x)=k x-k_3 x^3$ is used to represent the bistability. The modification creates a displacement range with the negative stiffness (a shadowed area in Fig.\,\ref{figure9}\,(a)). A bistable double-well potential $U(x)=- k x^2/2 + k_3 x^4/4 $ (Fig.\,\ref{figure9}\,(b)) visualises the bistability using the relationship between force and potential: $ f (x)=-\frac{\partial U(x)}{\partial x}$. The potential allows us to interpret cantilever motion as a particle moving along the potential profile. The local minima of the bistable profile (the solid line in Fig.\,\ref{figure9}\,(b)) correspond to two stable states, while a local maximum corresponds to an unstable state. The main motivation for exploring bistability is the expectation that switching between states efficiently expands the utilised frequency range of vibration that can be beneficial for broadband vibration. This motivation is more likely tightly linked to the phenomenon of stochastic resonance, which has significantly impacted the scientific community, shifting the paradigm from considering noise as a nuisance to an essential and nontrivial contribution in many situations, especially in living systems \cite{Gammaitoni:1998,Wellens:2004}. Unfortunately, the stochastic resonance is often misinterpreted (see \cite{McDonnell:2009,Khovanov:2021} for a discussion), leading many researchers to expect a synergy between bistability and noise in applied problems by default. However, as with many nonlinear effects, bistability can be beneficial only over some ranges of system and input parameters.

Let us remind ourselves that the cantilever nonlinearity is an internal, non-removable property. So, creating bistability is an example of modifying nonlinearity rather than converting a linear system into a nonlinear one.  

\subsubsection{ The response of the bistable harvester to a harmonic input }

Since the harmonic input $a(t)=A \sin (\omega t)$ is the de facto test signal used in engineering, let us consider the response of the harvester (\ref{eq_neh}) with bistable nonlinearity $f_s=0.5 (-x+x^3)$ to it. The other parameters of (\ref{eq_neh}) are the following: $\alpha=1$, $\psi=0.05$, $\lambda=1$ and $\kappa=0.5$. The natural frequency of the cantilever is equal to $1$, and the resonance frequency is around $\omega_{lr}\approx 0.9$ for the linearised harvester, which is a valid approximation of the nonlinear one for small $A$.  Select three frequencies: $\omega_1=0.9$, which is equal to the resonance one, $\omega_2=0.5$  and  $\omega_3=0.1$.  The output power, $P_v $, defined by the following expression 
\begin{align}
\label{pw2}
	P_v =  \frac{1}{\tau} \int_{t=0}^{t=\tau}  \dot{v}^2 dt       \  , 
\end{align} 
characterises the harvester performance. 

\begin{figure}[!tb]
\centering
	\includegraphics[height=0.33\columnwidth]{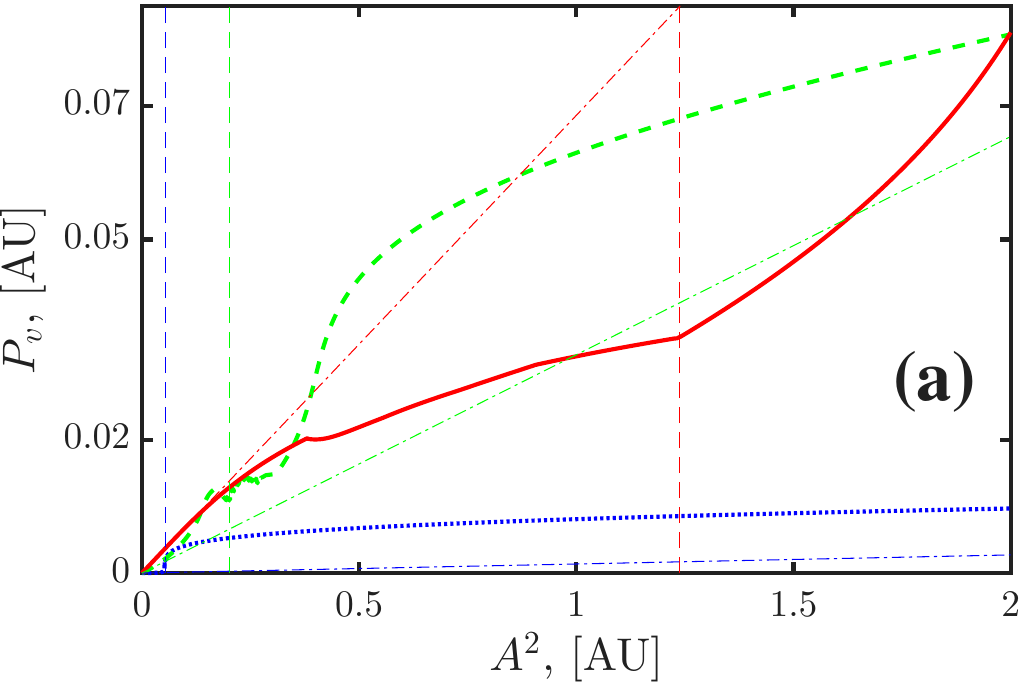} ~~
	\includegraphics[height=0.33\columnwidth]{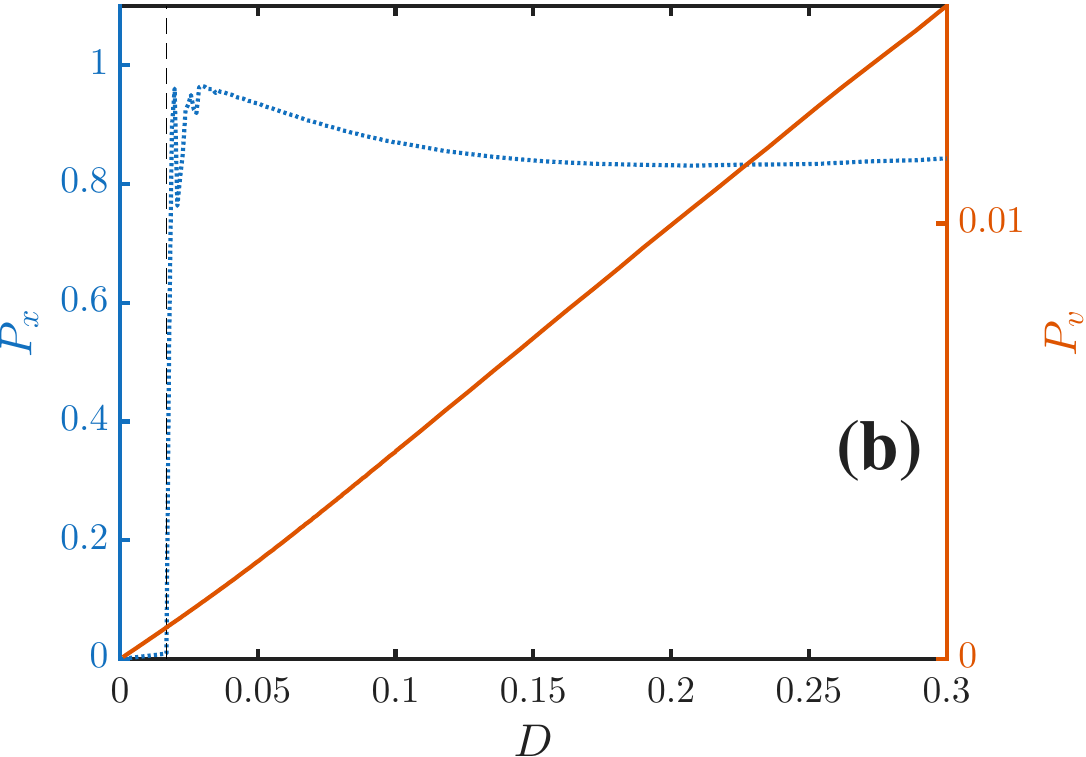}  
\caption{(a) The output power $P_v$ of the bistable harvester driven by the harmonic signal is shown for different frequencies: $\omega_1=0.9$ (the solid red line), $\omega_2=0.5$ (the dashed green line) and $\omega_3=0.1$ (the dotted blue line). The vertical thin dashed lines indicate the values of $A_{th}^2$ when the harvester starts to visit both states. The dashed-dotted lines correspond to the prediction of $P_v (A^2)$ assuming the linear response of the harvester. (b) The harvester output power $P_x$, estimated for the displacement, and $P_v$, estimated for voltage, are shown with the dotted and solid lines, respectively. Note that the y-axis for $P_x$ and $P_v$ are on the left and right sides of the figure, respectively. The vertical thin dashed line indicates the value of the noise intensity $D_{th}$ when the noise-induced transitions between the two states are started.  } \label{figure10}
\end{figure}

Figure~\ref{figure10}\,(a) shows the dependence of the power $P_v$ on  $A^2$ for selected frequencies. The thin vertical lines mark the threshold value $A_{th}$ at which the harvester begins to visit both states, i.e., when bistability becomes relevant.  The electrical part of the harvester, described by the second equation in (\ref{eq_neh}) for the piezoelectric layer, is a low-pass filter. Therefore, the value $A_{th}$ increases with the input frequency.  The thin dashed-dotted lines show the predicted power $P_v$ if the linear response was valid for the entire range of $A^2$. 

The most dramatic changes in the response are observed for low frequency $\omega_3$ (the dotted line in Fig.\,\ref{figure10}\,(a)). The harvester efficiency for small amplitude is poor since $\omega_3$ is far away from the resonance frequency $\omega_{lr}$. Around $A_{th}$, the power $P_v$  sharply increases, demonstrating that bistability improves the harvester's performance. Thus, if the harvester's natural frequency differs from the harmonic input frequency, introducing bistability can be an efficient strategy to improve energy harvesting.

Now let us consider the response for the frequency $\omega_1$ (the solid line in Fig.\,\ref{figure10}\,(a)), which is close to the resonance frequency $\omega_{lr}$.  Here, we can select two regions. One region corresponds to $A<A_{th}$, when oscillations occur around one of the states. For small amplitudes, the response is close to linear, and the conversion of vibration to electrical power is most efficient. Then there is a region where the nonlinearity reduces conversion efficiency because the resonance frequency shifts with increasing amplitude $A$.  In the second region, when $A>A_{th}$, the cantilever oscillates around both states and the rate of power growth is comparable with those observed for the linear response. Thus, modifying the nonlinearity by creating a bistable configuration improves energy conversion. A similar conclusion can be derived for $\omega_2$ (the dashed line in Fig.\,\ref{figure10}\,(a)), for which the conversion efficiency increases when $A^2>0.4$ when the cantilever starts to oscillate around two states. 

\subsubsection{ The response bistable harvester to broadband noise }

As mentioned above, bistability is considered a means to improve the harvester efficiency for broadband noise. For the ungrounded reason, it is expected that the high-frequency part of the noise induces switchings between the states, and therefore, the utilisation of broadband energy is observed. Unfortunately, this expectation is incorrect. The system's (harvester's) internal properties determine which part of the spectrum is used to induce the transition between the states. This spectrum part is only slightly wider than that observed in the frequency response obtained for the harmonic input. Therefore, a high-frequency part of the noise is irrelevant for the noise-induced transition. 

Another misleading narrative is based on the consideration of the cantilever displacement. Let's consider the harvester response to the white noise. In Figure \ref{figure10}\,(b), the powers $P_x$ and $P_v$ are shown as a function of the noise intensity $D$. The values $P_x$ were calculated using the expression (\ref{pw2}) where the displacement $x$ replaces the voltage $v$. The figure shows that there is indeed a sharp increase in $P_x$ around $D_{th}=0.017$ (the vertical dashed line) when the cantilever starts jumping between states. However, the power $P_v$, which reflects the harvester performance, does not demonstrate similar behaviour to $P_x$. In fact, the voltage $v$ is proportional to the cantilever velocity $\dot{x}$, which does not exhibit bistable behaviour. 

\subsubsection{ Narrowband  stochastic oscillation }

\begin{figure}[!tb]
\centering
	\includegraphics[width=0.48\columnwidth]{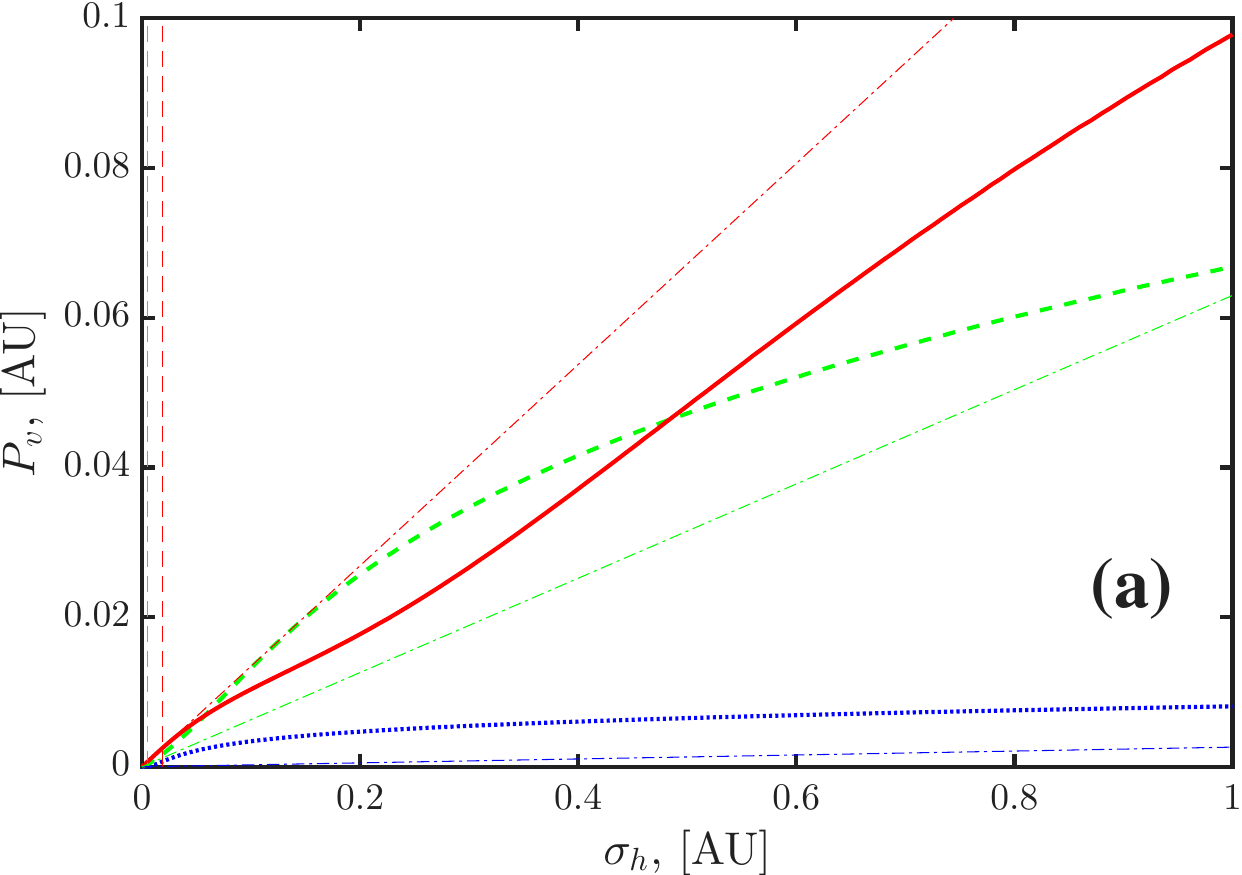} ~~
	\includegraphics[width=0.48\columnwidth]{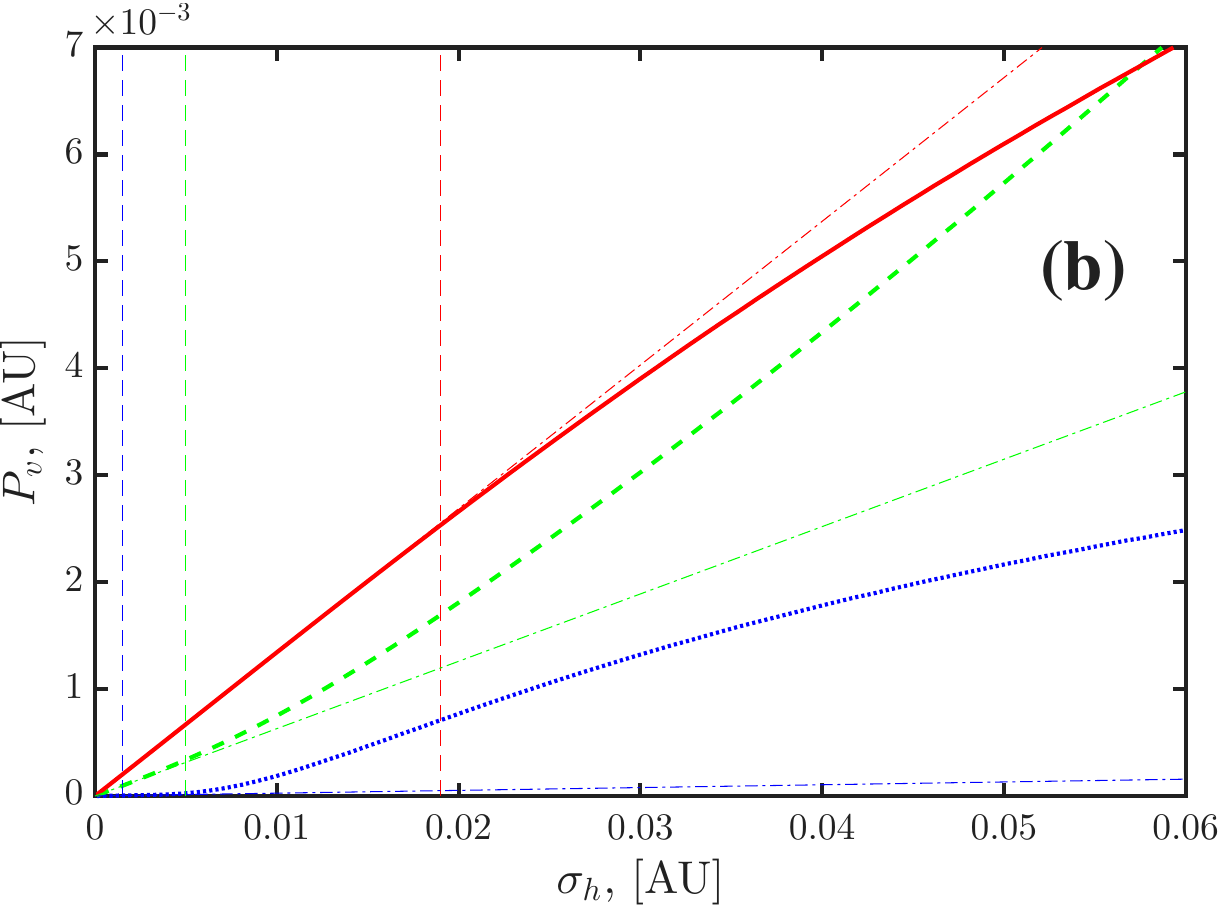}  
\caption{ (a) The output power $P_v$ of the bistable harvester driven by the harmonic noise is shown for different peak frequencies $\omega_p$: $\omega_1=0.9$ (the solid red line), $\omega_2=0.5$ (the dashed green line) and $\omega_3=0.1$ (the dotted blue line). The thin dashed lines indicate the noise intensity $\sigma_h$ when the noise-induced transitions between the two states are started.  The dashed-dotted lines correspond to the prediction of $P_v(\sigma_h)$ assuming the linear response of the harvester. (b) A zoomed-in part of Figure (a) for small noise intensity $\sigma_h$. } \label{figure11}
\end{figure}

Narrowband random vibration is possibly the ubiquitous type of ambient vibration of mechanical structures. Any structure has dimensions, which define vibration modes with specific frequencies. As a result, perturbations induce these narrowband modes. For example, broadband perturbations from the road result in narrowband spectral peaks in the suspension vibrations. Similar transformations are observed in civil structures \cite{Lin:04}, such as bridges, buildings, or floors. Often, the perturbations are random. Therefore, the test signal in the form of a single harmonic leads to irrelevant prediction, especially if the harvester demonstrates nonlinear properties. Another option is to use the white noise, which has a uniform spectral density, as a test signal. Since harmonic signals and white noise are used in many branches of science, there is a comprehensive literature on them. However, translating predictions from these widely applied test signals to the narrowband case is not straightforward. 

An adequate model of a narrowband noise is harmonic noise \cite{Schimansky:90,Dykman:91}, which is the output, $n_h(t)$, of a linear oscillator driven by white noise, $\xi(t)$. The harmonic noise is a solution of the following stochastic differential equation:
\begin{eqnarray}
\label{eq:hn}
\ddot n_h +\Gamma \dot{n_h}+\omega_h^2 n_h = \sqrt{D} \xi(t)  ,
\end{eqnarray}
where parameters $\Gamma$ and $\omega_h$ define the spectral density, $S_{h}(\omega)$, of the noise. The shape of $S_{h}(\omega)$ is close to the frequency response of the linear cantilever considered above (\ref{eq_linosc}), and it has the following form:
\begin{eqnarray}
\label{eq:sd_hn}
S_{h}(\omega) =\frac{D}{2\pi \left[\omega^2\Gamma^2+(\omega^2-\omega_h^2)^2 \right]} .
\end{eqnarray}
System (\ref{eq:hn}) is a linear filter for noise $\xi(t)$, and $n_h(t)$ has a Gaussian (normal) probability density similar to white noise $\xi(t)$. The output variable $n_h(t)$ has zero mean, $\langle n_h(t) \rangle =0$, and variance $\sigma_h=\langle n^2(t) \rangle = \frac{D}{\Gamma \omega_h^2}$ which defines the noise intensity. The maximum of $S_{h}(\omega)$ is located at frequency $\omega_p = \sqrt{\omega_h^2-\Gamma^2/2}$ and the spectral width is proportional to $\Gamma$. For small values of damping $\Gamma \ll \omega_h^2$, the spectrum of the harmonic noise becomes very narrow, approaching a delta-peak of the harmonic signal. However, the distributions of the harmonic noise and the harmonic signal differ.  In the opposite limit of a large damping, i.e. when $\Gamma \gg \omega_h^2$, the spectrum is flat, and the harmonic noise is close to and can be replaced with white noise. Note that the noise model to use depends heavily on the harvester's properties. If the frequency response of a harvester is significantly narrow, then the white noise is a relevant model. Harmonic noise should be used if its bandwidth is comparable to the harvester's frequency response bandwidth.

Let us consider the response of the bistable harvester to the harmonic noise and compare it to the response to the harmonic signal (Fig.\,\ref{figure10}\,(a)). Select (see \cite{Khovanov:2017} for details) the width of $S_{h}(\omega)$ close to the width of the frequency response of the linearised harvester, $\Delta \omega = 0.01$, and the location of spectral peaks at the same frequencies $\omega_1=0.9$, $\omega_2=0.5$ and $\omega_3=0.1$ as for the harmonic signals. The dependencies of the harvester power $P_v$ on the noise intensity $\sigma_h$ are shown in Figure \ref{figure11}. Similarly to Fig.\,\ref{figure10}, the vertical thin dashed lines mark the threshold value $\sigma_{h}^{th}$ when the noise is strong enough to induce jumping between the states, and the thin dashed-dotted lines show the prediction of the power $P_v$ based on the linearised harvester model. There are some differences in the dependences $P_v(\sigma_h)$ in comparison to the harmonic signal. For example, in the region $\sigma_h<\sigma_h^{th}$ (see Fig.\,\ref{figure11}\,(b)), the power $P_v(\sigma_h)$ linearly grows, and the power performance does not reduce as it is observed for the harmonic signal. 

The changes in bistable harvester performance are linked to switching rather than to the nonlinearity of each state. The rate of switchings between states grows with the noise intensities, and switchings are rare if the noise intensity is only slightly larger than $\sigma_h^{th}$. As a result, the frequent switchings and therefore, frequent oscillations between two states are only observed when $\sigma_h \gg \sigma_h^{th}$. So, bistability improves the harvester performance, but in a different way than for the harmonic signal. For the low frequency $\omega_p=\omega_3$, the switching between states results in a step-wise change in power $P_v(A^2)$ for the harmonic signal, and a similar change is more gradual in the harmonic noise case. 

This comparison shows that modelling predictions of nonlinear harvesters depend on the properties of ambient vibration. In work \cite{Khovanova:2014}, it has been shown that the efficiency of energy harvesting from narrowband vibrations strongly varies with cantilever nonlinearities and depends on the spectral width, $\Delta \omega$, and intensity, $\sigma_h$, of the harmonic noise.

Note that a comprehensive analysis of the harvester response to harmonic noise is a challenging task compared to the harmonic signal or white noise, for which the modelling predictions can be verified using similar modelling settings implemented in experiments using standardised shakers. The corresponding equipment is unavailable for the harmonic noise. Moreover, even for modelling purposes, there is no widely accepted framework similar to that used for the harmonic signal, where the frequency response is estimated over a range of frequencies.  Note that such a framework was proposed in work \cite{Khovanov:2017} to estimate the system response to harmonic noise by varying the locations of $\omega_p$, the spectral widths $\Delta \omega$, and the intensities $\sigma_h$. The value of the framework for the applied task remains unclear.

\section{ Conclusions }

A relatively straightforward task of energy harvesting, consisting of transferring mechanical energy to an electrical one, is quite complicated for ambient vibration. There are several reasons for the complications. From a technological point of view, a low power level associated with ambient vibration prevents the use of modern control engineering solutions that can provide sophisticated, adaptive pathways to facilitate and optimise energy production and conversion, because the solutions require energy. Therefore, optimised energy harvesting relies on properties of relatively simple components, such as a cantilever and a corresponding mechanical-electrical transformer, e.g., a piezoelectric layer. These components should be designed to maximise the output for a multi-facet input, which can be a mixture of several periodic and/or random components.

The examples of ambient vibrations considered show that their properties mismatch those of typical signals used in engineering design. The typical signal model is a harmonic signal, and sometimes it includes broadband noise with the normal distribution. While in the analysed examples, the harmonic components are present and the noise can be broadband, there are many harmonic components, some of which modulate the amplitudes of others. Also, the noise is a dominant contribution to ambient vibration. We can see strong broadband and narrowband noise with a non-normal, non-Gaussian probability density. Since a well-developed framework \cite{bendat_random_2011}, which is applied here to analyse examples of ambient vibration, is not consistently applied across the energy harvesting field, there is no comprehensive understanding of which signal models are more appropriate. Generic arguments linking a mechanical structure to a vibration source lead to the conclusion that narrowband noise is a typical representation of ambient vibration. In turn, the examples presented, however, show that this is not always the case. 

Another complication factor is the nonlinear behaviour of the mechanical part of the energy harvester, a cantilever, for a large deviation. Note that practically all engineering systems exhibit nonlinear behaviour. However, engineering design \cite{Thorby:2008} tends to utilise the system's parameters in a linear case, because it allows us to use a comprehensive set of available tools to predict and optimise the system for arbitrary inputs. In energy harvesting, the relative weakness of vibration implies maximising the cantilever deviation, which inevitably introduces nonlinearity into the analysis. Therefore, the traditional design approach requires a significant extension to account for various nonlinear effects. Roughly speaking, the traditional design \cite{Thorby:2008} is based on a single physical effect: linear resonance; the extension requires including many other effects, such as nonlinear resonance, parametric resonance, zero-dispersion resonance, stochastic resonance, and vibrational resonance. It is apparent that the terminology is very extensive in the nonlinear case. While many nonlinear effects are well understood and described, their routine use in engineering practice is absent, possibly because considering a particular nonlinear system is a fascinating task that can lead to unexpected, unique outputs, but requires numerical modelling with limited analytical tools, unlike the linear case, where the main tools are analytical.

Currently, energy harvesting is an active research area, with exploration of nonlinear design \cite{Jia:2020}. The main idea is to identify and implement modifications to the system's nonlinearity to maximise energy harvesting. The development of the nonlinear design is far from mature, as discussed in this manuscript; it has ups and downs. The strong noise in the ambient vibration is often overlooked and has a profound effect on the harvester's performance. It was shown that the response of nonlinear systems depends on both the properties of the nonlinearity and the external driving. The challenge is to combine and adjust these two factors to maximise the harvester performance. The bistability looks promising as a nonlinear modification. However, more research is needed, and a broader range of nonlinearities should be explored. Additionally, a more realistic model of ambient vibration should be embedded into the consideration. 

\section*{Disclosure statement}

The author reports there are no competing interests to declare. 

\section*{Funding}

No funding was obtained to support this work.


\small
\begin{thebibliography}{99}
\providecommand{\urlprefix}{Available from: }

\bibitem{Priya:09}
Priya~S, Inman~DE. Energy harvesting technologies. Berlin; Heidelberg:
  Springer-Verlag; 2009.

\bibitem{Beeby:11}
Kazmierski~T, Beeby~SE. Energy harvesting systems: Principles, modeling and
  applications. Science+Business Media, LLC: Springer; 2011.

\bibitem{Szarka:2018}
Szarka~GD, Stark~BH, Burrow~SG. Review of power conditioning for kinetic energy
  harvesting systems. IEEE Transactions on Power Electronics.
  2012;\hspace{0pt}27(2):803--815.

\bibitem{Qiu:06}
Qiu~H, Lee~J, Lin~J, et~al. Wavelet filter-based weak signature detection
  method and its application on rolling element bearing prognostics. Journal of
  Sound and Vibration. 2006;\hspace{0pt}289(4):1066--1090.

\bibitem{bendat_random_2011}
Bendat~JS, Piersol~AG. Random data: Analysis and measurement procedures. John
  Wiley \& Sons; 2011.

\bibitem{Shen:25}
Shen~Y, Zhang~X, Yang~L, et~al. A {VibV} dataset integrating vibration and
  vision for enhanced safety in self-driving tasks. Scientific Data.
  2026;\hspace{0pt}13(1):74.

\bibitem{car_data2}
{Menegazzo}~J, {von Wangenheim}~A. Multi-contextual and multi-aspect analysis
  for road surface type classification through inertial sensors and deep
  learning. In: 2020 X Brazilian Symposium on Computing Systems Engineering
  (SBESC); 2020. p. 1--8.

\bibitem{Kerschen:2006}
Kerschen~G, Worden~K, Vakakis~AF, et~al. Past, present and future of nonlinear
  system identification in structural dynamics. Mechanical Systems and Signal
  Processing. 2006;\hspace{0pt}20(3):505--592.

\bibitem{Thorby:2008}
Thorby~D. Structural dynamics and vibration in practice: An engineering
  handbook. Oxford: Butterworth-Heinemann; 2008.

\bibitem{Inman:2014}
Inman~DJ. Engineering vibration. 4th ed. Upper Saddle River, New Jersey:
  Pearson; 2014.

\bibitem{Lifshitz:2008}
Lifshitz~R, Cross~MC. Nonlinear dynamics of nanomechanical and micromechanical
  resonators. In: Schuster~HG, editor. Reviews of nonlinear dynamics and
  complexity. Vol.\,1; Chapter~1. Weinheim: Wiley-VCH; 2008. p. 1--52.

\bibitem{Kovacic:2011}
Kovacic~I, Brennan~MJE, editors. The Duffing equation: Nonlinear oscillators
  and their behaviour. Chichester: Wiley; 2011.

\bibitem{Butcher:16}
Butcher~JC. Numerical methods for ordinary differential equations. 3rd ed.
  Hoboken, New Jersey: John Wiley {\&} Sons; 2016.

\bibitem{Nayfeh:1981}
Nayfeh~AH. Introduction to perturbation techniques. New York: Wiley; 1981.

\bibitem{Moon:2004}
Moon~FC. Chaotic vibrations: An introduction for applied scientists and
  engineers. Hoboken, New Jersey: Wiley; 2004. 

\bibitem{Jia:2020}
Jia~Y. Review of nonlinear vibration energy harvesting: {Duffing}, bistability,
  parametric, stochastic and others. Journal of Intelligent Material Systems
  and Structures. 2020;\hspace{0pt}31(7):921--944.

\bibitem{Erturk:09}
Erturk~A, Hoffmann~J, Inman~DJ. A piezomagnetoelastic structure for broadband
  vibration energy harvesting. Applied Physics Letters.
  2009;\hspace{0pt}94(25):254102.

\bibitem{Litak:10}
Litak~G, Friswell~MI, Adhikari~S. Magnetopiezoelastic energy harvesting driven
  by random excitations. Applied Physics Letters.
  2010;\hspace{0pt}96(21):214103.

\bibitem{Khovanova:2014}
Khovanova~NA, Khovanov~IA, Davletzhanova~Z. Nonlinear energy harvesting from
  random narrow-band excitations. International Journal of Structural Stability
  and Dynamics. 2014;\hspace{0pt}14(8):1440026.

\bibitem{Mannella:Heun}
Mannella~R. Integration of stochastic differential equations on a computer.
  International Journal of Modern Physics C. 2002;\hspace{0pt}13(9):1177--1195.

\bibitem{Khovanova:12}
Khovanova~NA, Windelen~J. Minimal energy control of a nanoelectromechanical
  memory element. Applied Physics Letters. 2012;\hspace{0pt}101(2):024104.

\bibitem{Cottone:09}
Cottone~F, Vocca~H, Gammaitoni~L. Nonlinear energy harvesting. Physical Review
  Letters. 2009;\hspace{0pt}102(8):080601.

\bibitem{Gammaitoni:1998}
Gammaitoni~L, H{\"a}nggi~P, Jung~P, et~al. Stochastic resonance. Reviews of
  Modern Physics. 1998;\hspace{0pt}70(1):223--287.

\bibitem{Wellens:2004}
Wellens~T, Shatokhin~V, Buchleitner~A. Stochastic resonance. Reports on
  Progress in Physics. 2004;\hspace{0pt}67(1):45--105.

\bibitem{McDonnell:2009}
McDonnell~MD, Abbott~D. What is stochastic resonance? {Definitions},
  misconceptions, debates, and its relevance to biology. PLOS Computational
  Biology. 2009;\hspace{0pt}5(5):e1000348.

\bibitem{Khovanov:2021}
Khovanov~IA. The response of a bistable energy harvester to different
  excitations: the harvesting efficiency and links with stochastic and
  vibrational resonances. Philosophical Transactions of the Royal Society A:
  Mathematical, Physical and Engineering Sciences.
  2021;\hspace{0pt}379(2198):20200245.

\bibitem{Lin:04}
Lin~Y, Cai~G. Probabilistic structural dynamics: Advanced theory and
  applications. 2nd ed. New York; London: Mcgraw-Hill; 2004.

\bibitem{Schimansky:90}
{Schimansky-Geier}~L, Zulicke~C. Harmonic noise: Effect on bistable systems.
  Zeitschrift fur Physik B Condensed Matter. 1990;\hspace{0pt}79(3):451--460.

\bibitem{Dykman:91}
Dykman~MI, McClintock~PVE, Stein~ND, et~al. Quasimonochromatic noise: New
  features of fluctuations in noise-driven nonlinear systems. Phys Rev Lett.
  1991;\hspace{0pt}67:933--936.

\bibitem{Khovanov:2017}
Khovanov~IA, Khovanova~NA. Frequency response of an energy harvester to
  harmonic noise: Towards stochastic frequency response of nonlinear systems.
  In: 2017 International Conference on Noise and Fluctuations (ICNF). IEEE;
  2017.

\end{thebibliography}
\end{document}